\documentclass[12pt]{article}
\usepackage{amssymb,amsmath,amscd}
\usepackage{epsfig}
\usepackage{epstopdf}
\usepackage{latexsym}
\usepackage{graphicx}
\usepackage{booktabs}
\usepackage{bbm}
\usepackage[margin=15pt,small]{caption}
\usepackage{subcaption}
\usepackage{enumitem}
\usepackage{float}
\usepackage[toc]{appendix}
\usepackage[numbers,compress]{natbib}
\usepackage{tikz}
\usetikzlibrary{positioning}

\usepackage[T1]{fontenc}

\usepackage{color}
\usepackage{datetime}

\usepackage[
      colorlinks=false,
      linkcolor=darkblue,  
      urlcolor=blue,    
      filecolor=blue,     
      citecolor=red,
linktocpage=true,
      pdfstartview=FitV,
      bookmarksopen=true    
      ]{hyperref}

\setlist{nolistsep}

\ifpdf
\fi

\let\oldbibliography\thebibliography
\renewcommand{\thebibliography}[1]{\oldbibliography{#1}
\setlength{\itemsep}{0pt}} 

\newcommand*{\boxedcolor}{red}
\makeatletter
\renewcommand{\boxed}[1]{\textcolor{\boxedcolor}{%
  \fbox{\normalcolor\m@th$\displaystyle#1$}}}
\makeatother

\definecolor{cardinal}{rgb}{0.6,0,0}
\definecolor{darkgreen}{rgb}{0,0.5,0}
\definecolor{golden}{rgb}{0.92, 0.7, 0}
\definecolor{midnight}{rgb}{0, 0, 0.5}
\definecolor{darkblue}{rgb}{0.2, 0, 0.8}

\def\eql{~=~}

\def\cals#1{\mathcal{#1}}

\def\eql{=}

\newcommand{\fG}{\mathfrak{G}}
\newcommand{\fGt}{\widetilde{\mathfrak{G}}}

\newcommand{\zt}{\tilde{z}}
\newcommand{\fF}{\mathfrak{F}}

\newcommand{\fFt}{\widetilde{\mathfrak{F}}}
\newcommand{\fW}{\mathfrak{W}}
\newcommand{\fWt}{\widetilde{\mathfrak{W}}}

\newcommand{\at}{\tilde{a}}
\newcommand{\bt}{\tilde{b}}
\newcommand{\ct}{\tilde{c}}
\newcommand{\ma}{\mathfrak{a}}

\newcommand{\mat}{\tilde{\mathfrak{a}}}

\def\ruv{r_\text{UV}}

\def\SO{{\rm SO}}
\def\rU{{\rm U}}

\def\SU{{\rm SU}}

\def\fF{\frak F}
\def\fG{\frak G}
\def\fW{\frak W}

\def\cals#1{\mathcal{#1}}

\def\xiK{ }

\numberwithin{equation}{section} 

\begin{document}  

\begin{titlepage}
 
\medskip
\begin{center} 
{\Large \bf Mass Deformations of the ABJM Theory:\\[6 pt] The Holographic Free Energy}

\bigskip
\bigskip
\bigskip
\bigskip

{\bf Nikolay Bobev,${}^{(1)}$ Vincent S. Min,${}^{(1)}$ Krzysztof Pilch,${}^{(2)}$ and Felipe Rosso${}^{(2)}$  \\ }
\bigskip
${}^{(1)}$
Instituut voor Theoretische Fysica, KU Leuven,\\ 
Celestijnenlaan 200D, B-3001 Leuven, Belgium
\vskip 5mm
${}^{(2)}$ Department of Physics and Astronomy \\
University of Southern California \\
Los Angeles, CA 90089, USA  \\
\bigskip
\tt{nikolay.bobev,~vincent.min~@kuleuven.be; ~pilch,~felipero~@usc.edu}  \\
\end{center}

\bigskip
\bigskip
\bigskip
\bigskip

\begin{abstract}

\noindent  
\end{abstract}

\noindent We find a class of new supersymmetric Euclidean solutions in four-dimensional maximal gauged supergravity. The holographic dual description of these backgrounds is given by a mass-deformation of the ABJM theory with general values for the R-charges. We calculate the $S^3$ free energy for the supergravity backgrounds and find agreement with the supersymmetric localization calculation of the free energy in the large $N$ limit.

\vfill

\end{titlepage}


\newpage

\setcounter{tocdepth}{2}
\tableofcontents

\section{Introduction}

Supersymmetric localization is a powerful tool that allows an exact calculation of many observables in supersymmetric QFTs. In the context of holography, those  exact results can be viewed as robust predictions for the physics in the bulk.  This interplay between localization and holography has turned out particularly fruitful in the study of  three- and four-dimensional CFTs for which    many explicit examples of holographic duals are known. In particular, appropriate deformations of those CFTs have led to many new tests of the gauge/gravity duality away from conformality, see \cite{Freedman:2013ryh,Bobev:2013cja,Martelli:2011fu,Martelli:2011fw,Martelli:2012sz,Farquet:2014kma,Martelli:2013aqa,Bobev:2016nua,Gutperle:2018axv,Toldo:2017qsh,Bobev:2018ugk} for a non-exhaustive list of references. Our focus in this paper is  a specific realization of this general idea to study  certain deformations of the ABJM SCFT.  

The ABJM SCFT  \cite{Aharony:2008ug}, or simply ABJM for short, describes the low-energy dynamics on the world volume of $N$ \hbox{M2-branes}. It is an $\mathcal{N}=6$, $\rU(N)_k\times \rU(N)_{-k}$ Chern-Simons matter theory, which is conformal and  for $k=1,2$  enjoys an enhanced $\mathcal{N}=8$ supersymmetry. Our goal is to study this $\cals N=8$  SCFT theory on $S^3$ in the presence of scalar deformations that break the conformal symmetry, but still preserve $\mathcal{N}=2$ supersymmetry. There are two distinct types of such deformations. One is the so-called {real mass deformation}, which can be thought of as turning on a vacuum expectation value for the scalar in a background $\mathcal{N}=2$ vector multiplet that couples to a global flavor symmetry of the theory. The other type is a more standard mass deformation of the superpotential. 

In the presence of both  types of deformations,  it is possible to compute the free energy of the theory on $S^3$, defined as the logarithm of the partition function, as a function of the real and superpotential masses. The crucial observation here is that the path integral of the theory on $S^3$ localizes to a matrix integral \cite{Kapustin:2009kz}, which in  the  large $N$ limit can be evaluated explicitly using a saddle point approximation  \cite{Jafferis:2011zi}.   

The real masses can also be thought of as a general assignment  of   R-charges for the chiral superfields in the ABJM theory. To determine the values of these charges at the superconformal point, one can employ $F$-maximization \cite{Jafferis:2010un,Closset:2012vg,Pufu:2016zxm}. For the extremal values of the R-charges and for the vanishing superpotential mass, ABJM is in its conformal vacuum.

The $S^3$ free energy for general real and superpotential masses scales as $N^{3/2}$, which suggests that it should be possible to compute it via holography in string or M-theory. Indeed, a number of pertinent results are already available in the literature as we now summarize.

In the absence of any deformation, the holographic dual of ABJM is  given by the well-known AdS$_4\times S^7$ solution of eleven-dimensional supergravity. More precisely, one should analytically continue AdS$_4$ to the hyperbolic space, $\mathbb{H}^4$, with an $S^3$ boundary. Adding  a superpotential mass for one of the chiral superfields of ABJM results in an RG flow to an interacting  CFT in the IR with $\mathcal{N}=2$ supersymmetry and $\SU(3)$ flavor symmetry. This SCFT, henceforth referred to as  mABJM,  was studied in \cite{Benna:2008zy,Klebanov:2008vq,Jafferis:2011zi} from the QFT perspective and in \cite{Warner:1983vz,Corrado:2001nv,Ahn:2000mf,Ahn:2000aq,Bobev:2009ms,Bobev:2018uxk} holographically. Both for ABJM and mABJM, the large $N$ calculation of the $S^3$ free energy agrees with the regularized on-shell action of the corresponding bulk AdS$_4$ (or $\mathbb{H}^4$) solution \cite{Drukker:2010nc,Jafferis:2011zi}. 
For vanishing superpotential masses,  the ABJM SCFT on $S^3$ deformed by real mass terms, that is with an arbitrary R-charge assignment, has its supersymmetry broken to $\mathcal{N}=2$. These deformations induce  RG flows, which are holographically dual to smooth Euclidean supergravity solutions constructed in \cite{Freedman:2013ryh}. The smooth cap off in the bulk of the supergravity solutions in \cite{Freedman:2013ryh} is a manifestation of the IR cutoff provided by the finite radius of $S^3$. 

In this paper we find  gravitational solutions that correspond to the deformation of the ABJM theory with  both non-trivial superpotential and real masses turned on.
The superpotential and real mass deformations of interest here are associated with operators in the $\mathcal{N}=8$ energy momentum multiplet of the ABJM SCFT. These operators are dual to the scalar fields in the four-dimensional $\mathcal{N}=8$ $\SO(8)$ gauged supergravity \cite{deWit:1982bul}, which in turn is a consistent truncation of eleven-dimensional supergravity on $S^7$ \cite{deWit:1986oxb,Nicolai:2011cy}. For that reason we will construct our solutions  within the four-dimensional theory.

The supergravity calculation involves several steps. We start by  identifying a suitable consistent truncation of the four-dimensional $\mathcal{N}=8$ $\SO(8)$ gauged supergravity. As we discuss in some detail below,  this truncation turns out to be  precisely the one obtained in \cite{Bobev:2018uxk} to study certain supersymmetric AdS$_4$ black holes. Since the deformations we are turning on  are scalar operators in ABJM on $S^3$,   we are looking for  solutions that preserve the isometry of $S^3$ in the bulk and have an $S^3$ boundary. To this end, we   study in detail the supersymmetry variations in the truncation of  the four-dimensional $\mathcal{N}=8$ theory, analytically continue them to the Euclidean signature, and then derive a set of first order ordinary differential equations (the Euclidean BPS equations  compatible with the second order equations of motion) for the metric and scalar fields that depend on the radial holographic coordinate only. 
Our BPS equations admit regular solutions, analogous to the ones in \cite{Freedman:2013ryh}, for which 
the $S^3$ collapses smoothly in the bulk. We construct these solutions explicitly through analytic and numerical techniques. 

The  non-trivial gravitational backgrounds that we find  are dual to the RG flows in the ABJM theory triggered by the presence of the non-trivial real masses as well as the superpotential mass parameter. In addition to these holographic RG flows, our supergravity truncation contains also two supersymmetric AdS$_4$ vacua. In one of them all scalars vanish and  the full $\SO(8)$ gauge symmetry of the supergravity theory is preserved. This is the dual of the conformal vacuum of the ABJM theory. The other vacuum has non-vanishing constant scalar fields that preserve an $\SU(3)\times \rU(1)$ subgroup of $\SO(8)$. This background is  dual to the $\mathcal{N}=2$ mABJM SCFT discussed in \cite{Benna:2008zy,Klebanov:2008vq,Jafferis:2011zi,Bobev:2018uxk}.

To calculate the partition function for this class of supergravity solutions, we must carefully apply the holographic renormalization formalism \cite{Skenderis:2002wp}. A subtlety here  is that in order to preserve supersymmetry we have to add a particular finite counterterm to the usual divergent counterterms that render the bulk on-shell action finite. An additional subtlety is the well-known fact that the proper treatment of the ABJM theory in a holographic setup requires the  alternative quantization of the scalars and the standard quantization of the pseudoscalars in the four-dimensional $\mathcal{N}=8$ supergravity  \cite{BREITENLOHNER1982249,Klebanov:1999tb}. Both of these subtleties have arisen also in previous studies of the holographic description of the ABJM theory \cite{Freedman:2013ryh,Freedman:2016yue}.\footnote{The presence of finite counterterms in holographic renormalization is also encountered in the five-dimensional Euclidean supergravity solutions discussed in \cite{Bobev:2013cja,Bobev:2016nua}.} At the end, we show that the on-shell action evaluated on our supergravity solutions precisely agrees with the ABJM partition function on $S^3$ obtained using supersymmetric localization. 

We continue in the next section  with a short summary of the ABJM theory and its  deformations, and the known results for the large $N$ limit of the corresponding partition functions on $S^3$. In Section~\ref{sec:Sugra}, we discuss the  consistent truncation  that captures the deformations of the ABJM theory of interest and the corresponding   Euclidean BPS equations. In Section~\ref{sec:SolBPS} we obtain the required solutions by a mixture of analytic and numerical methods. We carefully carry out the holographic renormalization of the on-shell action and implement the proper quantization of the scalar fields by a Legendre transform in Section \ref{sec:HoloRen}. The result shows the  perfect agreement between the partition functions of the holographic duals. We conclude in Section~\ref{sec:Conclusions} with a summary and some comments. The three appendices are devoted to important technical details. In Appendices~\ref{appendixA} and \ref{AppendixB}, we derive the supersymmetry variations and the BPS equations in the Euclidean regime and then, in Appendix~\ref{AppendixC}, the Euclidean equations of motion.

\section{Field theory}
\label{sec:CFT}

In this section, following \cite{Freedman:2013ryh} and \cite{Bobev:2018uxk}, we present a short summary of the ABJM theory, its relevant deformations and the results from supersymmetric localization of interest here. The ABJM SCFT \cite{Aharony:2008ug} is an $\rU(N)_k\times \rU(N)_{-k}$ Chern-Simons matter theory with $\mathcal{N}=6$ supersymmetry. The theory can be formulated in $\mathcal{N}=2$ superspace in terms of two vector multiplets, four chiral multiplets, $A_a$, $B_c$ with $a,c=1,2$ and superpotential
\begin{equation}\label{eqn:superpotABJM}
W \sim \text{Tr}\left(\epsilon^{ab}\epsilon^{cd}A_aB_cA_bB_d\right)\,.
\end{equation}
For $k=1,2$ there is an enhancement of supersymmetry to $\mathcal{N}=8$ and the R-symmetry is $\SO(8)$. From now on we focus on the case $k=1$. Imposing that the R-charge of the superpotential in \eqref{eqn:superpotABJM} is equal to two, leads to the following constraint on the R-charges of the chiral superfields
\begin{equation}\label{DeltaconstrABJM}
\Delta_{A_1}+\Delta_{A_2}+\Delta_{B_1}+\Delta_{B_2}=2\;.
\end{equation}
The SO(8) R-symmetry of the conformal theory then leads to
\begin{equation}\label{conformalABJM}
\Delta_{A_1}=\Delta_{A_2}=\Delta_{B_1}=\Delta_{B_2}=\frac{1}{2} \;.
\end{equation}
Another way to find these values for the R-charges is to use F-maximization \cite{Jafferis:2010un}. To this end one needs to compute the free energy of the theory on $S^3$ for arbitrary values of the R-charges and maximize the resulting function. This can be done using supersymmetric localization \cite{Kapustin:2009kz} and in the large $N$ limit the result reads \cite{Jafferis:2011zi}
\begin{equation}\label{FS3}
F_{S^3}=\frac{4\sqrt{2}\pi}{3}N^{3/2}\sqrt{\Delta_{A_1}\Delta_{A_2}\Delta_{B_1}\Delta_{B_2}} \;.
\end{equation}
Indeed, taking into account the constraint \eqref{DeltaconstrABJM}, we find that \eqref{FS3} is maximized at the values of the R-charges in \eqref{conformalABJM}.
For values of the R-charges different from the ones in \eqref{conformalABJM}, but still obeying the constraint \eqref{DeltaconstrABJM}, the theory has $\mathcal{N}=2$ supersymmetry but is no longer conformal. As explained in detail in \cite{Freedman:2013ryh}, this can also be understood as a result of deforming ABJM by coupling it to background vector multiplets. To understand this better, parametrize the solutions to the constraint \eqref{DeltaconstrABJM} as
\begin{equation}\begin{aligned}\label{deltai}
\Delta_{A_1} &= \frac{1}{2} +\delta_1+\delta_2+\delta_3 \,, \quad & \Delta_{A_2} &= \frac{1}{2} +\delta_1-\delta_2-\delta_3 \;, \\
\Delta_{B_1} &= \frac{1}{2} -\delta_1+\delta_2-\delta_3 \,, \quad & \Delta_{B_2} &= \frac{1}{2} -\delta_1-\delta_2+\delta_3 \;.
\end{aligned}\end{equation}
The parameters $\delta_{1,2,3}$ can then be thought of as the values of the complex scalars, which reside in the background $\mathcal{N}=2$ abelian vector multiplets that couple to the maximal torus of the $\SU(4)$ flavor symmetry of the ABJM theory. These parameters are often referred to as real masses.

In addition, the ABJM theory admits a more standard mass term given by deforming the superpotential in \eqref{eqn:superpotABJM} by
\begin{equation}\label{CPWsuperp}
\Delta W=m \text{Tr}(T^{(1)}A_1)^2\,.
\end{equation}
Here $T^{(1)}$ is a monopole operator which has vanishing R-charge, see \cite{Jafferis:2011zi,Freedman:2013ryh}. This superpotential deformation breaks the flavor symmetry of the model from $\SU(4)$ to $\SU(3)$ and triggers an RG flow to an interacting $\mathcal{N}=2$ SCFT in the IR. This theory was studied in \cite{Benna:2008zy,Klebanov:2008vq,Jafferis:2011zi} and was referred to as mABJM in \cite{Bobev:2018uxk}. Given that the superpotential in \eqref{CPWsuperp} has R-charge two, we immediately find that 
\begin{equation}\label{DeltaconstrmABJM}
\Delta_{A_1} = 1\,, \qquad \Delta_{A_2}+\Delta_{B_1}+\Delta_{B_2}=1 \,.
\end{equation}
The large $N$ limit of the $S^3$ free energy of this deformation of the ABJM theory was computed in \cite{Jafferis:2011zi} and reads 
\begin{equation}\label{FS3CPWgen}
F_{S^3} =\frac{4\sqrt{2}\pi}{3}N^{3/2}\sqrt{\Delta_{A_2}\Delta_{B_1}\Delta_{B_2}} \;.
\end{equation}
This result amounts to simply implementing \eqref{DeltaconstrmABJM} in \eqref{FS3} and agrees with the intuition that the chiral superfield $A_1$ is integrated out from the dynamics at low energies. Applying \hbox{F-maximization} to the expression in \eqref{FS3CPWgen}, subject to \eqref{DeltaconstrmABJM}, we find that the superconformal \hbox{R-charges} of the mABJM SCFT are
\begin{equation}\label{conformalmABJM}
\Delta_{A_2}=\Delta_{B_1}=\Delta_{B_2}=\frac{1}{3} \;.
\end{equation}
The same result follows from the $\SU(3)$ symmetry.

The free energy in \eqref{FS3CPWgen} makes sense for general values of the R-charges, which can again be interpreted as real masses. Taking into account the constraint in \eqref{DeltaconstrmABJM}, combined with \eqref{deltai}, leads to
\begin{equation}\label{deltaconstr}
\delta_1+\delta_2+\delta_3 = \frac{1}{2} \,,
\end{equation}
or, alternatively,
\begin{equation}\label{Deltadelta}
\Delta_{A_2}=2\delta_1\,, \qquad \Delta_{B_1}=2\delta_2\,, \qquad \Delta_{B_2}=2\delta_3 \,.
\end{equation}
Therefore, we have two independent real mass parameters in addition to the superpotential mass~$m$. This is compatible with the fact that the $\SU(4)$ flavor symmetry of the ABJM theory is broken  by the superpotential in \eqref{CPWsuperp} to $\SU(3)$, which has a two-dimensional maximal torus. The main goal of the following sections is to derive the free energy in \eqref{FS3CPWgen} using holography by explicitly constructing supergravity solutions, which encode the superpotential and real mass deformations, and evaluating their on-shell action. 

Before we embark on this task, let us emphasize that the free energy in \eqref{FS3CPWgen} is independent of the dimensionless parameter, $m R_{S^3}$, where $m$ is the superpotential mass and $R_{S^3}$ is the radius of the sphere.\footnote{We set $R_{S^3}=1$ throughout this paper.} This is due to the fact that in the supersymmetric localization calculations the path integrals of the ABJM and mABJM theories depend only the real masses $\delta_i$. Thus, while the dependence of the free energy on the parameters $\delta_i$ is continuous, the role of the parameter $m$ is simply to impose the constraint \eqref{DeltaconstrmABJM}, or equivalently \eqref{deltaconstr}. Therefore the superpotential mass $m$ in this setup can be viewed as a discrete parameter, which changes the free energy from \eqref{FS3} for $m=0$ to \eqref{FS3CPWgen} for $m\neq 0$. A somewhat singular limit is obtained by taking $m R_{S^3} \to \infty$. Then the theory is effectively in flat space and the IR cutoff provided by the finite radius of $S^3$ is removed. The RG flow can then reach the strongly interacting mABJM SCFT and the R-charges are fixed to their superconformal values in \eqref{conformalmABJM}. This simple picture of the RG flows triggered by the real and superpotential mass deformations is confirmed by the supergravity solutions studied below. 

\section{The supergravity model}
\label{sec:Sugra}

The deformation of the ABJM theory by real and superpotential masses discussed above preserves a $\rU(1)^3$ subgroup of the $\SO(8)$ R-symmetry and is triggered by operators in the energy momentum tensor multiplet of the theory. To construct supergravity solutions dual to this deformation, one thus needs to consider a $\rU(1)^3$-invariant truncation of the maximal ${\rm SO(8)}$ gauged supergravity \cite{deWit:1982bul} and construct asymptotically AdS$_4$ supersymmetric Euclidean solutions of this model. Precisely such a truncation has been constructed recently in \cite{Bobev:2018uxk}. To obtain it, one first considers the fields of the gauged supergravity theory invariant under the $\rU(1)^2$ maximal torus of the $\SU(3)$ flavor symmetry of the mABJM theory. As discussed in detail in \cite{Bobev:2018uxk}, this leads to a $\mathcal{N}=2$ gauged supergravity theory with three Abelian vector multiplets and one hypermultiplet. The bosonic fields of this model consist of the metric, four vector fields and five complex scalars. Imposing an additional $\rU(1)$ symmetry, dual to the supeconformal R-symmetry of the mABJM SCFT, leads to a further consistent truncation containing all four vector fields, the three complex scalar fields, $z_i$, in the vector multiplets, but only a single hyperscalar, $z$,  which is  a complex scalar field from the hypermultiplet. Here we are interested in solutions of this supergravity model that preserve the isometries of $S^3$ and have  the four Abelian gauge fields consistently set to zero.\footnote{See the comment below \eqref{Kgflds}.} Note that setting the hyperscalar, $z$, to zero yields the well-known STU model of four-dimensional gauged supergravity used in the construction of \cite{Freedman:2013ryh}. We continue with the salient features of the action and BPS equations for the resulting model with four complex scalar fields. Further details are discussed in Appendix \ref{appendixA} and \cite{Bobev:2018uxk}. 

\subsection{The Euclidean action}
\label{EucAct}

The Euclidean bulk action 
\begin{equation}\label{bulkS}
S_\text{bulk}\eql\int d^4x \sqrt{\det g_{\mu\nu}}\,\cals L_\text{bulk}\,,
\end{equation}
is obtained by performing the Wick rotation on the bosonic Lagrangian \eqref{4dLagrLor} of the truncated theory. The resulting bulk Lagrangian is\footnote{The length scale, $L$, is related to the gauge coupling constant, $g$, by $L=1/\sqrt 2 g$. We have also fixed the four-dimensional Newton constant by setting $8 \pi G_{(4)}=1$.}
\begin{equation}\label{4dLagr}
 \mathcal{L}_\text{bulk} = - \frac{1}{2} R + \sum_{i=1}^3\frac{g^{\mu\nu}\partial_\mu z_i \partial_\nu \zt_i}{(1-z_i \zt_i)^2} + \frac{g^{\mu\nu}\partial_\mu z \partial_\nu \zt}{(1-z \zt)^2} + {1\over 2L^2} \,\mathcal{P} \,,
\end{equation}
where  the Euclidean metric, $g_{\mu\nu}$, is real and positive definite. As usual,  in the Euclidean regime the complex scalar fields and their complex conjugates should be treated as  independent fields. Following  \cite{Freedman:2013ryh}, we denote the latter by   $\tilde z_i$ and $\tilde z$  rather than $\bar z_i$ and $\bar z$, respectively. Since the Wick rotation does not act on the manifold parametrized by the scalars, all eight scalars, $z_i$, $\tilde z_i$, $z$ and $\tilde z$, still take values in the Poincar\'e disk and hence are complex with modulus less than one.

The Euclidean scalar potential, $\cals P$, obtained from \eqref{scpoten} is complex  in general and can be written in the following form,
\begin{equation}\label{P}
\mathcal{P} = \frac{1}{2} \left(\sum_{i=1}^3\fF_i \fFt_i+\frac{4 z\zt }{(1-z \zt)^2}\, \fG \,\fGt - 3 \,\fW \,\fWt \right) \,,
\end{equation}
where, cf.\ \eqref{defofW} and  \eqref{defofG}, 
\begin{equation}\label{WFG}
\begin{split}
\fW &= e^{K_V/2} \frac{1}{1- z \zt}  \left[\,2\,(z_1 z_2 z_3 -1)+z \zt \,(1-z_1)(1-z_2)(1-z_3)  \, \right] \,, \\[6 pt]
\fF_i &= e^{K_V/2} \frac{1}{1- z \zt} \left[\, 2\left(\zt_i - \frac{z_1 z_2 z_3}{z_i}\right) +z \zt \,\frac{1-\zt_i}{1-z_i}(1-z_1)(1-z_2)(1-z_3) \,\right]\,, \\[6 pt]
\fG &= e^{K_V/2}\big[\,2(z_1z_2z_3-1)+(1-z_1)(1-z_2)(1-z_3)\,\big] \,, \\
\end{split}
\end{equation}
with the tilded functions  obtained by the exchange $z_i\leftrightarrow \zt_i$ and $z\leftrightarrow \zt$. The  K\"ahler potential  in \eqref{WFG}, cf.\ \eqref{KvP}, is
\begin{equation}\label{}
e^{K_V/2}  = {1\over (1-z_1 \zt_1)^{1/2}(1-z_2 \zt_2)^{1/2}(1-z_3 \zt_3) ^{1/2}}\,.
\end{equation}
The functions defined in \eqref{WFG} satisfy 
\begin{equation}\begin{aligned}\label{iden}
\fF_i &= -(1-z_i \zt_i)\left[\,\frac{\partial}{\partial z_i}\fW+\frac{1}{2}\Big(\frac{\partial}{\partial z_i}K_V\Big)\,\fW\,\right] \,, & \qquad  \fG &= (1-z \zt)^2 \frac{\partial}{\partial(z \zt)}\fW \,, \\
\end{aligned}\end{equation}
with similar relations for $\widetilde\fF_i$ and $\widetilde \fG$.

\subsection{The BPS equations}
\label{BPSeqs}

Following the general procedure for the holographic calculation of the partition function, $F_{S^3}$, proposed in \cite{Freedman:2013ryh} and  summarized  in Section~\ref{sec:HoloRen}, we are interested in  supersymmetric solutions to the equations of motion for the action \eqref{bulkS} in which the metric  has the $S^3$-sliced form,  
\begin{equation}\label{Ansatz}
ds^2 =  L^2 e^{2A} ds^2_{S^3} +e^{2B} dr^2\,,
\end{equation}
and the metric functions, $A$ and $B$, as well as the scalar fields, $z_i$, $\tilde z_i$, $z$, and $\tilde z$,  depend  only on the radial 
coordinate, $r$. Such solutions are obtained by solving the Euclidean BPS equations that follow from the vanishing of the Wick rotated supersymmetry variations of the fermion fields in the Lorentzian $\cals N=2$ supergravity. 

In Appendix~\ref{AppendixB} we show that those BPS equations can be reduced to the following ``flow'' equations for the scalars and the metric function:\footnote{We set $\xi=-1$ in \eqref{solforM}-\eqref{zzeqs}.}  
\begin{equation}\label{zzjEqs}
z_j'\eql -{e^B\over 2L}\,\xiK\,(1-z_j\tilde z_j)\,{\fG^{1/2}\over\widetilde\fG^{1/2}}\,\widetilde\fF_j\,,\qquad
\tilde z_j'\eql -{e^B\over 2L}\,\xiK\,(1-z_j\tilde z_j)\,{\widetilde \fG^{1/2}\over \fG^{1/2}}\,\fF_j\,,
\end{equation}
\begin{equation}\label{zzEqs}
{z'\over z}\eql {e^B\over L}\,\xiK \,\fG^{1/2}\,\widetilde \fG^{1/2}\,,\qquad 
{\tilde z'\over \tilde z}\eql {e^B\over L}\,\xiK\,\fG^{1/2}\,\widetilde \fG^{1/2}\,,
\end{equation}
and 
\begin{equation}\label{dAeqss} 
A'\eql {e^B\over L}\,\Big[\,\pm  e^{-A}-{1\over 2}\, \xiK\,{\widetilde\fG^{1/2}\over \fG^{1/2}}\, \fW \,\Big]\eql {e^B\over L}\,\Big[ \,\mp e^{-A}-{1\over 2}\,\xiK\,{\fG^{1/2}\over\widetilde \fG^{1/2}}\,\widetilde\fW \,\Big]
\,,
\end{equation}
where prime denotes the derivative with respect to the radial coordinate, $r$.

The two signs in  \eqref{dAeqss} correspond to two branches, I and II, of solutions related by the exchange of tilded and untilded fields. By combining the two equations  in \eqref{dAeqss} one arrives at the perhaps more familiar looking flow equation,
\begin{equation}\label{AsqEqs}  
(A')^2\eql {1\over L^2}\Big[e^{-2(A-B)}+{1\over 4}e^{2B}\,\fW\,\widetilde\fW\,\Big]\,,
\end{equation}
which is the same for both branches. One can also solve \eqref{dAeqss}
to obtain the metric function in terms of the scalars,
\begin{equation}\label{expAsol}
e^{A}\eql \mp 4\,{\fG^{1/2}\,\widetilde \fG^{1/2}\over \fG\,\widetilde\fW-\widetilde\fG\,\fW}\,.
\end{equation}
Then  \eqref{AsqEqs}  follows from \eqref{expAsol} using the BPS equations \eqref{zzjEqs} and \eqref{zzEqs} for the scalars.
Similarly, the BPS equations \eqref{zzjEqs}-\eqref{dAeqss} imply the equations of motion  given in Appendix~\ref{AppendixC}. 

The BPS equations \eqref{zzjEqs}-\eqref{dAeqss} have been derived assuming that the scalar fields are not constant.  For constant scalars,  one is simply left with the single flow equation \eqref{AsqEqs} and the algebraic constraints \eqref{alge1}-\eqref{alge2}, which fix the scalars to their critical values.

In simplifying our BPS equations we have also assumed nontrivial hyperscalars, $z$ and $\tilde{z}$. This means that in the limit when the hyperscalars are turned off, \eqref{zzjEqs} and \eqref{dAeqss} become  equivalent to the BPS equations in \cite{Freedman:2013ryh}, but with an additional subtlety, see Section~\ref{ssec:AnSol}  and  Appendix~\ref{AppendixB}.  Our parametrization of the scalars, $z_i$ and $\tilde z_i$, is related to the one in \cite{Freedman:2013ryh} by 
\begin{equation}\label{FBzs}
z_i^\text{here} = -\zt_i^\text{FP}\,,\qquad  \zt_i^\text{here} = -z_i^\text{FP}\,,
\end{equation}
and is the same as in \cite{Bobev:2018uxk}. 

Note that the metric function $B$ remains undetermined, neither by the BPS equations \eqref{zzjEqs}-\eqref{dAeqss}  nor the equations of motion \eqref{EOMs}, but it can be removed by a suitable reparametrization of the radial coordinate. However, keeping it explicit allows for a quick transition between different gauges.  
The two gauges that we will be using in the following are: the conformal gauge,
\begin{equation}\label{conf}
e^{2B} = \frac{L^2}{r^2} e^{2A} \qquad \Longrightarrow \qquad ds^2 = \frac{L^2}{r^2} e^{2A}\left(dr^2 + r^2 ds^2_{S^3}\right) \;,
\end{equation}
and the Fefferman-Graham (FG) gauge, 
\begin{equation}\label{FG}
e^{2B} = L^2 \qquad \Longrightarrow \qquad ds^2 = L^2 \left( d\rho^2 + e^{2A} ds^2_{S^3}\right) \;,
\end{equation}
where  the radial coordinate will be denoted by  $\rho$ to distinguish it from the radial coordinate, $r$, in the conformal gauge.

\section{Solutions to the BPS equations}
\label{sec:SolBPS}

\subsection{Two  AdS solutions}
\label{ssec:2ads}

As a warm-up exercise let us consider two $\mathbb{H}^4$ (i.e.  Euclidean AdS$_4$)  solutions corresponding to two supersymmetric critical points of the potential \eqref{P}:
\begin{align}
\text{SO(8)}: & & z_i&= \zt_i = 0 \;, &  z &= \zt = 0\;, &  \mathcal{P}_* &= -6\;, \label{SO8} \\
\text{W}: & & z_i &= \zt_i = \sqrt 3-2\;, & z\zt &= {1\over 3}\,, & \mathcal{P}_* &= -\frac{9\,\sqrt{3}}{2} \label{W} \;,
\end{align}
Here  $\cals P_*$ is the value of the potential at the critical point. Note that the W-point exists for any choice of $z$ and $\zt$ obeying \eqref{W} so in fact we have a one-parameter family of critical points.\footnote{In \cite{Bobev:2018uxk} the Lorentzian signature W-point is at $z= - \bar{z}=\pm\frac{i}{\sqrt{3}}$.}  

For both points, we only need to solve the   BPS equation \eqref{AsqEqs} for the metric,  which now reduces to
\begin{equation}\label{}
(A')^2\eql {e^{2B}\over L^2}\,\Big(e^{-2A}-{1\over 6}\,\cals P_*\Big)\,.
\end{equation}
Its solution in the   conformal gauge \eqref{conf} yields the  metric,
\begin{equation}\label{H4metric}
ds^2\eql {4L^2 \over (1-r^2/\ruv^2)^2}\,(dr^2+r^2\,ds_{S^3}^2)\,,\qquad \ruv^2\eql -{6\over \cals P_*}\,,
\end{equation}
with the constant curvature radius $L_*=\ruv L$. 

We have chosen to normalize the radial coordinate, $r$, in \eqref{H4metric} in a somewhat nonstandard way, which will turn out to be convenient when solving the general BPS equations below. In this parametrization, the $S^3$ boundary in the ``UV region'' is  at $r\to \ruv$. The metric then caps off smoothly in the  ``IR region''   at $r\to r_\text{IR}=0$, where 
\begin{equation}\label{expAir}
e^{2A}\eql 4r^2+\cals O(r^4)\,,
\end{equation}
universally for both solutions and the metric approaches that of flat space.

The $\SO(8)$ point in \eqref{SO8} is dual to the $\mathcal{N}=8$ conformal vacuum of the ABJM theory. The W-point preserves $\SU(3)\times\rU(1)$ of the $\SO(8)$ gauge symmetry and was first found in \cite{Warner:1983vz}. It is the gravitational dual to the conformal vacuum of the $\mathcal{N}=2$ mABJM SCFT.

\subsection{The asymptotic analysis}
\label{ssec:asymlis}

To get a general picture of the space of solutions to the BPS equations \eqref{zzjEqs}-\eqref{dAeqss}, we first perform the standard near-boundary UV expansion  and then derive  conditions that a regular solution  must satisfy in the bulk in the IR limit.

\subsubsection{The UV asymptotics}
\label{sec:UV}

We are interested in solutions that asymptote to an $\mathbb H^4$ solution with vanishing scalar fields in the UV. Using the conformal gauge \eqref{conf} for the metric, the asymptotic expansions of the scalar fields obtained by solving the equations of motion are given by:
\begin{equation}\label{ruv1}
\begin{split}
z_i(r)& \eql \alpha_i(r-\ruv)+\beta_i(r-\ruv)^2+\ldots\,,\\[6 pt]
\tilde z_i(r) & \eql \tilde \alpha_i(r-\ruv)+\tilde \beta_i(r-\ruv)^2+\ldots\,,\\[6 pt]
z(r) & \eql \alpha\,(r-\ruv) +\beta (r-\ruv)^2+\ldots\,,\\[6 pt]
\tilde z(r) & \eql \tilde \alpha\,(r-\ruv)+\tilde \beta (r-\ruv)^2+\ldots\,.\\[6 pt]
\end{split}
\end{equation}
Note that by rescaling the radial coordinate we could set the UV radius $\ruv=1$. However, it is more convenient to keep it here explicitly as to allow for a universal normalization of the radial coordinate in the IR region as in \eqref{expAir}. 

Substituting the expansions \eqref{ruv1} into the BPS equations \eqref{zzjEqs} and \eqref{zzEqs}, we find the following constraint on the leading order parameters,
\begin{equation}\label{alind}
\alpha_1+\alpha_2+\alpha_3-\tilde \alpha_1-\tilde \alpha_2-\tilde \alpha_3\eql -{2\over \ruv}\,.
\end{equation}
The next order expansion determines the subleading coefficients that are given by 
\begin{gather}\label{ruv2}
\beta_i\eql {\alpha\tilde\alpha\over 2}-{3\over 2}{\alpha_i\over \ruv}-{\tilde\alpha_1\tilde\alpha_2\tilde \alpha_3\over\alpha_i}\,,\qquad \tilde\beta_i\eql {\alpha\tilde\alpha\over 2}+{1\over 2}{\tilde \alpha_i\over \ruv}-{\alpha_1\alpha_2\alpha_3\over\alpha_i}\,,
\\[6 pt]
\beta\eql \alpha\,\Big(\alpha_1+\alpha_2+\alpha_3+{1\over 2\ruv}\Big)\,,\qquad \tilde \beta\eql  \tilde \alpha\,\Big(\tilde \alpha_1+\tilde \alpha_2+\tilde \alpha_3-{3\over 2\ruv}\Big)\,,
\end{gather}
and similarly the higher order terms in the expansions  \eqref{ruv1} are determined by the leading coefficients. 
The resulting expansion of the metric function can be found from \eqref{expAsol},
\begin{equation}\label{ruv3}
\begin{split}
e^{2A}& \eql {\ruv^2\over (r-\ruv)^2}+{\ruv\over r-\ruv}-\Big[\,{1\over 4}+{1\over 3}\ruv^2(\alpha_1\tilde \alpha_1+\alpha_2\tilde\alpha_2+\alpha_3\tilde\alpha_3)\Big]
\\[6 pt]
& \quad +\ruv^2\,\Big[ (\alpha_1\alpha_2\alpha_3+\tilde \alpha_1\tilde \alpha_2\tilde \alpha_3)-{\alpha\tilde\alpha\over 2}(\alpha_1+\alpha_2+\alpha_3+
\tilde \alpha_1+\tilde \alpha_2+\tilde \alpha_3)\Big](r-\ruv)+\ldots\,.
\end{split}
\end{equation}

One can also perform the asymptotic analysis above in the FG-gauge \eqref{FG}, which is perhaps more familiar in the context of holographic renormalization, see \cite{Skenderis:2002wp} for a review. The asymptotic coefficients in this gauge are more directly related to the field theory quantities that we are interested in and a comparison with some results in \cite{Freedman:2013ryh} that we would like to use  is more straightforward. 

In the UV region, the  standard FG radial coordinate, $\rho$, is given by
\begin{equation}\label{UVexp}
{r\over \ruv}\eql 1-2\,e^{-\rho}+2\,e^{-2\rho}+\ldots\,.
\end{equation}
In the UV limit ($\rho\to\infty$),  the scalar fields, including both hyperscalars, $z$ and $\tilde z$, have the following expansions:
\begin{equation}\label{FGexpand}
\begin{split}
z_i(\rho)= a_i e^{-\rho}+b_i e^{-2\rho} &+  \dots  \,,  \qquad \zt_i(\rho) = \at_i e^{-\rho} + \bt_i e^{-2\rho}+\dots \,,
\\[6 pt]
z(\rho)= a e^{-\rho} + b e^{-2\rho}&  +  \dots  \,, \qquad  \zt(\rho)  = \at e^{-\rho} + \bt e^{-2\rho} +\dots\,,
\end{split}
\end{equation}
where the expansion coefficients in \eqref{FGexpand} are related to those in \eqref{ruv1} by
\begin{gather}\label{FGtoconf}
a_i\eql -2\ruv \alpha_i\,,\qquad \tilde a_i\eql  -2\ruv \tilde \alpha_i\,,\\[6 pt]
b_i\eql 4\ruv^2 \beta_i+2\ruv \alpha_i\,,\qquad \tilde b_i\eql 4\ruv^2 \tilde \beta_i+2\ruv \tilde \alpha_i\,,
\end{gather}
and
\begin{gather}\label{}
a\eql - {2\ruv\alpha}\,,\qquad \tilde a\eql - {2\ruv\tilde \alpha}\,,\\[6 pt]
 b\eql 4\ruv^2\beta+2\ruv\alpha\,,\qquad \tilde b\eql 4\ruv^2\tilde \beta+2\ruv\tilde \alpha\,.
\end{gather}
For later use, let us also write the identities corresponding to \eqref{alind} and \eqref{ruv2},
\begin{equation}\label{aconstr}
a_1 + a_2 + a_3 -\at_1 - \at_2 - \at_3   = 4 \,,
\end{equation}
and
\begin{equation}\label{abrels1}\begin{split}
b_i &= \frac{1}{2}\left[\,a \at + a_i (a_1+a_2+a_3-\at_1-\at_2-\at_3)-2 \frac{\at_1 \at_2 \at_3}{\at_i}\,\right]\,, \\[6 pt]
\bt_i &= \frac{1}{2}\left[\,a \at - \at_i (a_1+a_2+a_3-\at_1-\at_2-\at_3)-2 \frac{a_1 a_2 a_3}{a_i}\,\right]\,, 
\end{split}\end{equation}
\begin{equation}\label{abrels2}
\begin{split}
b &= \frac{a}{2}\left(a_1+a_2+a_3+\at_1+\at_2+\at_3\right) \,,\\
\bt &= \frac{\at}{2}\left(a_1+a_2+a_3+\at_1+\at_2+\at_3\right) \,,
\end{split}
\end{equation}
respectively. 

Finally, let us define
\begin{equation}\label{Ddefs}
\Delta_i\eql {1\over 4}(a_i-\tilde a_i)\,,
\end{equation}
in terms of which  \eqref{aconstr} becomes 
\begin{equation}\label{Delid}
\Delta_1+\Delta_2+\Delta_3\eql 1\,.
\end{equation}
The relation \eqref{Delid} is the supergravity analog of the constraint between the real masses in the mABJM field theory \eqref{DeltaconstrmABJM}. 
Note that in the absence of the hyperscalars we would not have the relation in \eqref{aconstr}.

\subsubsection{The IR asymptotics}
\label{ss:IRasym}

It is clear from the form of the metric \eqref{conf} in the conformal gauge that regular solutions must cap off at $r=0$ where the sphere, $S^3$, shrinks to zero. This means that the metric becomes flat, see \eqref{expAir},
\begin{equation}\label{Aat0}
A(r)=A_0+\log r+\cals O(r^2)\,,
\end{equation}
and  the scalars have finite values,
\begin{equation}\label{irinitv}
z_i(0)\eql c_i\,,\qquad \tilde z_i(0)\eql \tilde c_i \,,\qquad z(0)\eql c\,,\qquad \tilde z(0)\eql \tilde c\,,
\end{equation}
and finite (vanishing) derivatives. Then \eqref{zzjEqs} and \eqref{zzEqs} imply that $\fG \,\widetilde \fF_i$, $ \widetilde \fG\,\fF_i$ and $\fG\,\widetilde \fG$  
must   vanish at $r=0$. Substituting \eqref{Aat0} in \eqref{dAeqss} and expanding to the leading order, we note that the $1/r$ pole cancels in only one equation in each pair. The cancellation of the $1/r$ pole  in the other equation requires that $\widetilde \fG(r)=\cals O(r^2)$ for the branch I and $\fG(r)=\cals O(r^2)$ for the branch II. This yields the following boundary conditions for regular solutions at $r=0$:
\begin{equation}\label{Branches}
\text{I.}\qquad \widetilde \fF_i(0)\eql0\,,\quad \widetilde \fG(0)\eql 0 \,;\qquad\qquad
\text{II.} \qquad  \fF_i(0)\eql0\,,\quad  \fG(0)\eql 0 \,.
\end{equation}

Since the two branches are related by the exchange of tilded and untilded fields, in the following we will consider only the first branch. Solving the equations in \eqref{Branches} we find that the constants $c_i$ are determined by $\tilde c_i$ and $x_0\equiv c\tilde c$,
\begin{equation}\label{solci}
c_i\eql {2\,\tilde c_j\tilde c_k-x_0(1-\tilde c_j)(1-\tilde c_k)\over 2-x_0(1-\tilde c_j)(1-\tilde c_k)}\,,\qquad \text{$(ijk)$-cyclic}\,,
\end{equation}
where $\tilde c_i$ satisfy a cubic constraint that follows from $\widetilde \fG(0)=0$, cf.\ \eqref{WFG},
\begin{equation}\label{cubic}
2(\tilde c_1\tilde c_2\tilde c_3-1)+(1-\tilde c_1)(1-\tilde c_2)(1-\tilde c_3)\eql 0\,.
\end{equation}
In the next section we show that these  conditions completely specify regular solutions modulo the rescaling of the radial coordinate, which is fixed universally for all solutions by  imposing \eqref{expAir}. 

It is worth stressing that the  cubic constraint \eqref{cubic} is a consequence of the coupling to the hypermultiplet scalars. Indeed, the corresponding derivation of the IR asymptotics with only vector scalars in  \cite{Freedman:2013ryh} yields as expected \eqref{solci} with $x_0=0$, but unconstrained  constants $\tilde c_i$.

\subsubsection{Summary}

The UV asymptotic analysis yields the parameters, $\alpha_i$, $\tilde \alpha_i$ and  $\alpha$, $\tilde\alpha$ or, equivalently, $a_i$, $\tilde a_i$ and  $a$, $\tilde a$  subject to the constraint \eqref{alind} and \eqref{aconstr}, respectively.  Hence in the UV region we have a 7-parameter family of  asymptotic solutions to the BPS equations. The question is  which of those extend to regular solutions in the bulk, where the asymptotic expansion in the IR region yields only a 4-parameter family of solutions  parametrized by $c_i$, $\tilde c_i$, $c$ and $\ct$ that must satisfy \eqref{solci} and \eqref{cubic}.
Our task now is to determine how these two families of asymptotic solutions are related. It appears that the only way to answer this question is to solve the BPS equations explicitly. Unfortunately, apart from a couple of special cases, this can  be done only numerically.

\subsection{Analytic solutions in the limit of vanishing hyperscalars}
\label{ssec:AnSol}

In the limit of vanishing hyperscalars, $z$ and $\tilde z$, the BPS equations  \eqref{zzjEqs}-\eqref{dAeqss}, modulo  the cubic constraint \eqref{cubic} in the IR, are equivalent to the BPS equations in \cite{Freedman:2013ryh}, which can be solved in closed analytic form. Using the field redefinition \eqref{FBzs}, we obtain an explicit family of solutions to our equations given by  
\begin{equation}\label{FPsol}
z_i(r)\eql c_i\,f(r)\,,\qquad \tilde z_i(r)\eql \tilde c_i\,f(r)\,,
\end{equation}
where\footnote{Note that the radial coordinate in \cite{Freedman:2013ryh} is rescaled such that $\ruv^\text{FP}=1$.}
\begin{equation}\label{theff}
f(r)\eql {1-r^2/r_\text{UV}^2\over 1-\tilde c_1\tilde c_2\tilde c_3\,r^2/r_\text{UV}^2}\,,\qquad r_\text{UV}^2\eql  1-\tilde c_1\tilde c_2\tilde c_3\,.
\end{equation}
From \eqref{solci} with $x_0=0$ we have
\begin{equation}\label{}
c_i\eql {\tilde c_1\tilde c_2\tilde c_3\over\tilde c_i}\,,
\end{equation}
where $\tilde c_i$ are constrained by \eqref{cubic}. Substituting the solution \eqref{FPsol} in \eqref{expAsol} and then using  \eqref{cubic} repeatedly to simplify the expression, we obtain
\begin{equation}\label{expAan}
e^{2A}\eql 4r^2\,{(1-\tilde c_1\tilde c_2\tilde c_3 \,r^4/\ruv^4)\over (1-r^2/\ruv^2)^2(1-\tilde c_1\tilde c_2\tilde c_3 \, r^2/\ruv^2)^2}\,,
\end{equation}
which agrees with \cite{Freedman:2013ryh}. We emphasize here that for $x_0=0$ the hyperscalars $z$ and $\tilde{z}$ do not flow and strictly speaking we are not forced to impose the cubic constraint in \eqref{cubic}. Nevertheless we do so because we are ultimately interested in the space of solutions to the BPS equations for which $z$ and $\tilde{z}$ are non-trivial. 

From the explicit solution \eqref{FPsol}-\eqref{theff}, we can  read off the UV asymptotics. After using  \eqref{ruv1} and \eqref{FGtoconf} in \eqref{FPsol}, we find:
\begin{equation}\label{fpais}
a_i\equiv -2\,\ruv\, z_i'(\ruv)\eql {4\,c_i\over 1-\tilde c_1\tilde c_2\tilde c_3}\,,\qquad \tilde a_i\equiv  -2\,\ruv \,\tilde z_i'(\ruv)\eql {4 \,\tilde c_i\over 1-\tilde c_1\tilde c_2\tilde c_3}\,.
\end{equation}
Substituting \eqref{fpais}  in \eqref{Ddefs}, we obtain the following map:
 \begin{equation}\label{solDel}
\Delta_1\eql {(1+\tilde c_2)(1+\tilde c_3)\over (1-\tilde c_2)(1-\tilde c_3)}\,,\qquad 
\Delta_2\eql {(1+\tilde c_3)(1+\tilde c_1)\over (1-\tilde c_3)(1-\tilde c_1)}\,,\qquad
\Delta_3\eql {(1+\tilde c_1)(1+\tilde c_2)\over (1-\tilde c_1)(1-\tilde c_2)}\,,
\end{equation}
between the IR and the UV data for this class of solutions. In fact, in the analysis in \cite{Freedman:2013ryh}, the analogous map was crucial for establishing the equality of partition functions on both sides of the correspondence. Hence, our  task here is to understand whether and if so how  \eqref{fpais} and \eqref{solDel} are modified when $x_0\neq 0$ and the hypermultiplet scalar has a non-trivial profile.

It was observed   previously in \cite{Bobev:2018uxk} that the map \eqref{solDel}  provides a  ``linearization'' of the cubic constraint in the sense that \eqref{Delid} holds if and only if \eqref{cubic} is satisfied.  Before we proceed, let us discuss briefly  the ranges of the parameters on both sides of this map.

For real $\tilde c_i$, with $|\tilde c_i|<1$, satisfying \eqref{cubic}, the corresponding $\Delta_i$ lie in the interval $0<\Delta_i<1$ and satisfy  \eqref{Delid}. The map is in fact one-to-one, which can be verified by solving \eqref{solDel} for the $\tilde c_i$'s. The solutions are double-valued, but only one branch lies within the unit disks. For complex $\tilde c_i$, we should impose an additional constraint that a solution has a good asymptotically $\mathbb{H}^4$ UV region. Given $\ruv$ in \eqref{theff}, we must set $0\leq \tilde c_1\tilde c_2\tilde c_3<1$. 
We have checked numerically,  that for such complex $\tilde c_i$ we have $0\leq \text{Re}\,\Delta_i\leq 1$, but have not found a simple characterization of the allowed region for the imaginary parts of $\Delta_i$'s. It appears that the restriction of the real parts of $\Delta_i$'s to the $[0,1]$ interval depends crucially on the reality of the product $\tilde c_1\tilde c_2\tilde c_3$.

\subsection{Numerical solutions of the BPS equations}
\label{ssec:NumSol}

In this section we present a numerical evidence for the existence of regular solutions with nontrivial hypermultiplet scalars and then study the resulting mapping between the IR and the UV data.  To this end let us first simplify further the set of equations that we need to solve.

The equality between the right hand sides of the two flow equations for the hyperscalars, $z$ and $\tilde z$, in \eqref{zzEqs} implies that the ratio  $z/\tilde z$ must be constant.
The two hyperscalars  enter  the BPS equation  \eqref{zzjEqs} and  \eqref{dAeqss} only through the product $z\tilde z$ in $\fF_i$, $\widetilde \fF_i$, $\fW$ and $\widetilde \fW$, while $\fG$ and $\widetilde\fG$ do not depend on them. Furthermore, in the solution \eqref{expAsol} the terms with $z\tilde z$ cancel out.  This means that all equations  are invariant under the constant rescaling $(z,\tilde z)\to (\lambda z,\lambda^{-1}z)$ and we can set the  ratio $z/\tilde z$  to any constant value.  This is reflected also by the fact that we can consistently rewrite the BPS equations and the equations of motion in terms of the composite field,\footnote{One may recover all the equations with $z$ and $\tilde z$ using \eqref{Xtozz} and setting $z=\lambda\sqrt X$, $\tilde z\eql \lambda^{-1}\sqrt X$ where $\lambda$ is a constant.}
\begin{equation}\label{Xtozz}
X\eql z\tilde z\,.
\end{equation}
This field in general can be complex  with   $|X|<1$.

It will be convenient to work in the  conformal gauge, where we can use \eqref{expAsol} to eliminate the metric functions from the other BPS equations, which leaves us with the following set of flow equations that involve only scalar fields: 
\begin{equation}\label{zzeqsn}
\begin{split}
r\,{dz _i\over dr} & \eql {2}\,(1-z_i\tilde z_i)\,{\fG\over \fG\,\fWt-\fGt\,\fW}\,\widetilde \fF_i\,,\qquad
r\,{d\tilde z_i\over dr}  \eql {2}\,(1-z_i\tilde z_i)\,{\fGt\over \fG\,\fWt-\fGt\,\fW}\, \fF_i\,,
\end{split}
\end{equation}
\begin{equation}\label{Xeqsn}
r\,{dX\over dr}\eql -{8}\,X\,{\fG\,\fGt\over \fG\,\fWt-\fGt\,\fW}\,.
\end{equation}

Note that all square-roots in \eqref{zzjEqs} and \eqref{zzEqs} have cancelled out and the right hand sides in the  equations above are rational functions of the scalars. With the boundary conditions in Section~\ref{ss:IRasym}, the IR point $r=0$ is then a regular singular point of the first order system  \eqref{zzeqsn}-\eqref{Xeqsn} whose solution can be expanded into  a  power series that converges as long as the right hand sides of the equations remain analytic (see, e.g. \cite{Fukuhara:1961a}).

It is instructive to see how this works for the explicit solution in Section~\ref{ssec:AnSol}. The power series for the function $f(r)$  in  \eqref{theff} around $r=0$ converges for 
\begin{equation}\label{}
r^2< R^2\qquad \text{where}\qquad R^2\eql {\ruv^2\over \tilde c_1\tilde c_2\tilde c_3}>\ruv^2\,,
\end{equation}
and hence, quite remarkably, the solution for the scalar fields is analytic in the entire region between the IR and the UV, and in fact well beyond it. It is only the metric function \eqref{expAan} that diverges at $r=\ruv$.

Turning on the hypermultiplet scalars  modifies  the IR boundary conditions in  \eqref{solci} through $x_0=X(0)$ and  adds the flow equation \eqref{Xeqsn} for $X(r)$. However, it does not modify in any way the analytic properties of the equations. It is thus reasonable to expect that, at least  for $x_0$ small enough, the power series solution that exists in the vicinity of $r=0$ should converge all the way through $r=\ruv$. To see how this expectation bears out in practice, we turn to numerical explorations. 

\subsubsection{Solutions in the symmetric sector}
\label{ss:su3}

The flow equations \eqref{zzeqsn} and \eqref{Xeqsn} can be consistently restricted to the subsector in which 
\begin{equation}\label{}
z_1\eql z_2\eql z_3\equiv \zeta\,,\qquad \tilde z_1\eql \tilde z_2\eql \tilde z_3\equiv \tilde \zeta, 
\end{equation}
with $\zeta$,  $\tilde \zeta $ and $X$ then satisfying,
\begin{equation}\label{zetaeqs}
\begin{split}
r\,{d\zeta\over dr} & \eql -{1\over 3}(1+4\zeta+\zeta^2)\,{2(\zeta-\tilde\zeta^2)+X(1-\zeta)(1-\tilde\zeta)^2\over (1-X)(\zeta-\tilde \zeta)(1-\tilde \zeta)}\,,\\[6 pt]
r\,{d\tilde \zeta\over dr} & \eql {1\over 3}(1+4\tilde \zeta+\tilde \zeta^2){2(\zeta^2-\tilde \zeta)-X(1-\zeta)^2(1-\tilde\zeta)\over (1-X)(\zeta-\tilde \zeta)(1-\tilde \zeta) }\,,\\[6 pt]
r\,{dX\over dr} & \eql -{4\over 3} \,X\,{(1+4\zeta+\zeta^2)\,(1+4\tilde \zeta+\tilde \zeta^2)\over (\zeta-\tilde\zeta)(1-\zeta\tilde\zeta)}\,.
\end{split}
\end{equation}
Imposing the IR boundary conditions \eqref{solci} and \eqref{cubic}, we find  
\begin{equation}\label{}
\zeta_0\equiv \zeta(0)\eql \frac{3 x_0 \left(3 x_0-4\right)+1}{\left(3 x_0-\sqrt{3}-2\right){}^2}\,,\qquad \tilde\zeta_0\equiv\tilde\zeta(0)\eql -2+\sqrt 3\,,\qquad x_0\equiv X(0)\,,
\end{equation}
that is we are left with one  free parameter, $x_0$. 
These conditions lead to a consistent recurrence for the series expansion provided we fix the scaling symmetry of the radial coordinate. This is done by requiring that the metric function given by \eqref{expAsol} has the leading term in the series expansion normalized as in \eqref{expAir}.

\begin{figure}[t]
\begin{center}
\includegraphics[height=1.65 in]{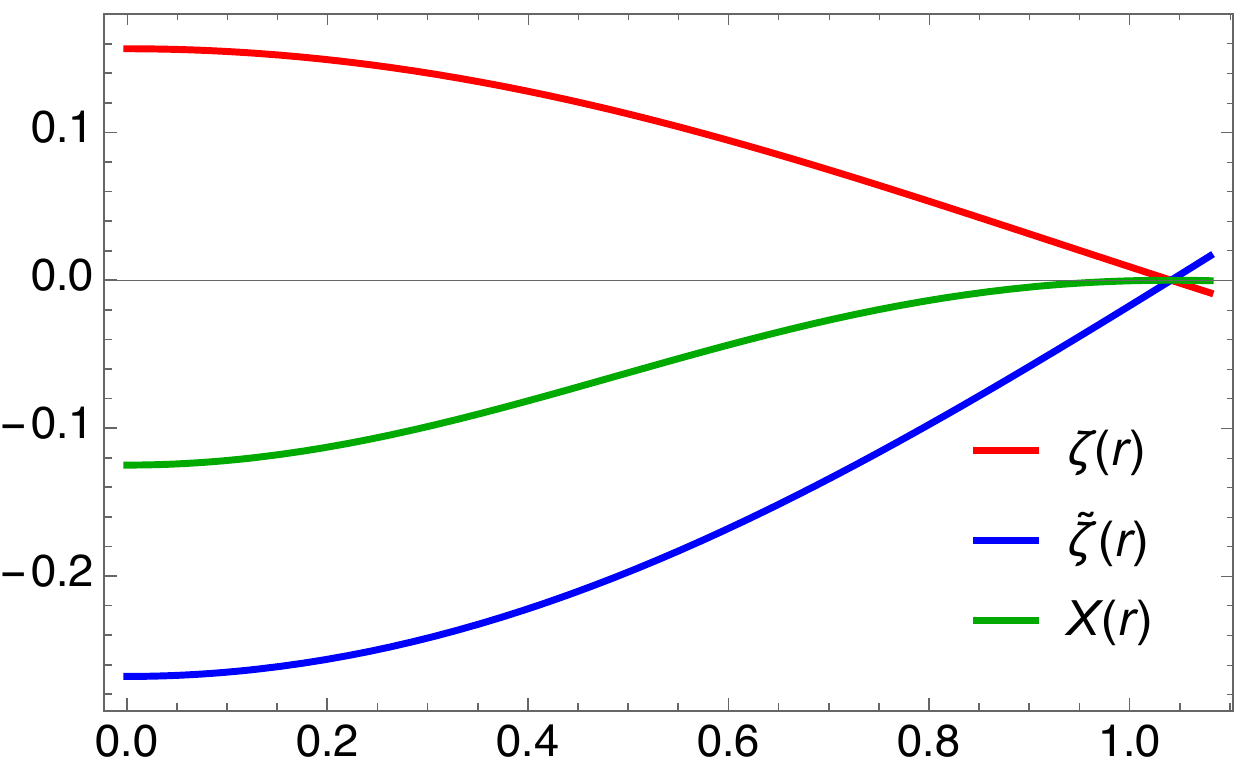}\qquad\qquad \includegraphics[height=1.65 in]{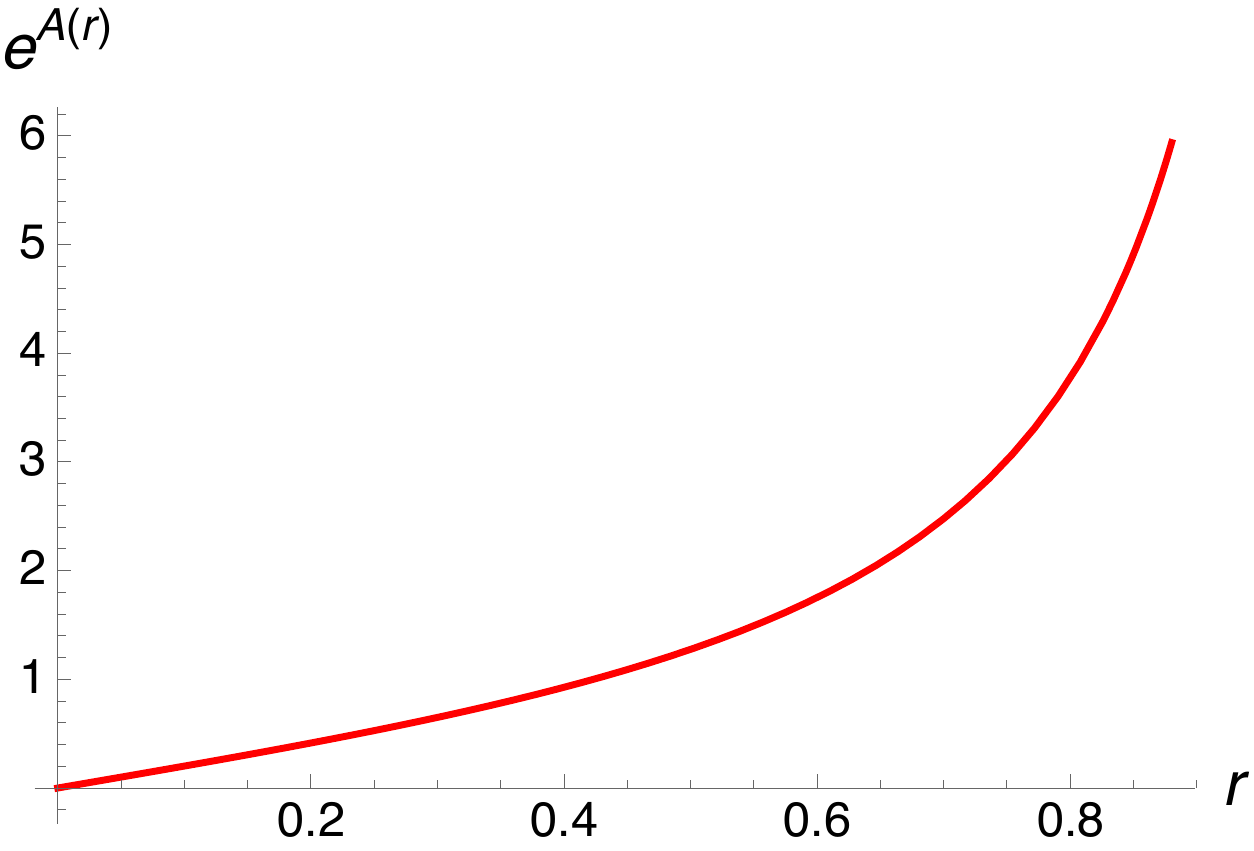}
\caption{Solution for $x_0=-0.125$ with $\ruv=1.041$.}
\label{fig1}
\end{center}
\end{figure}

\begin{figure}[t]
\begin{center}
\includegraphics[height=1.75 in]{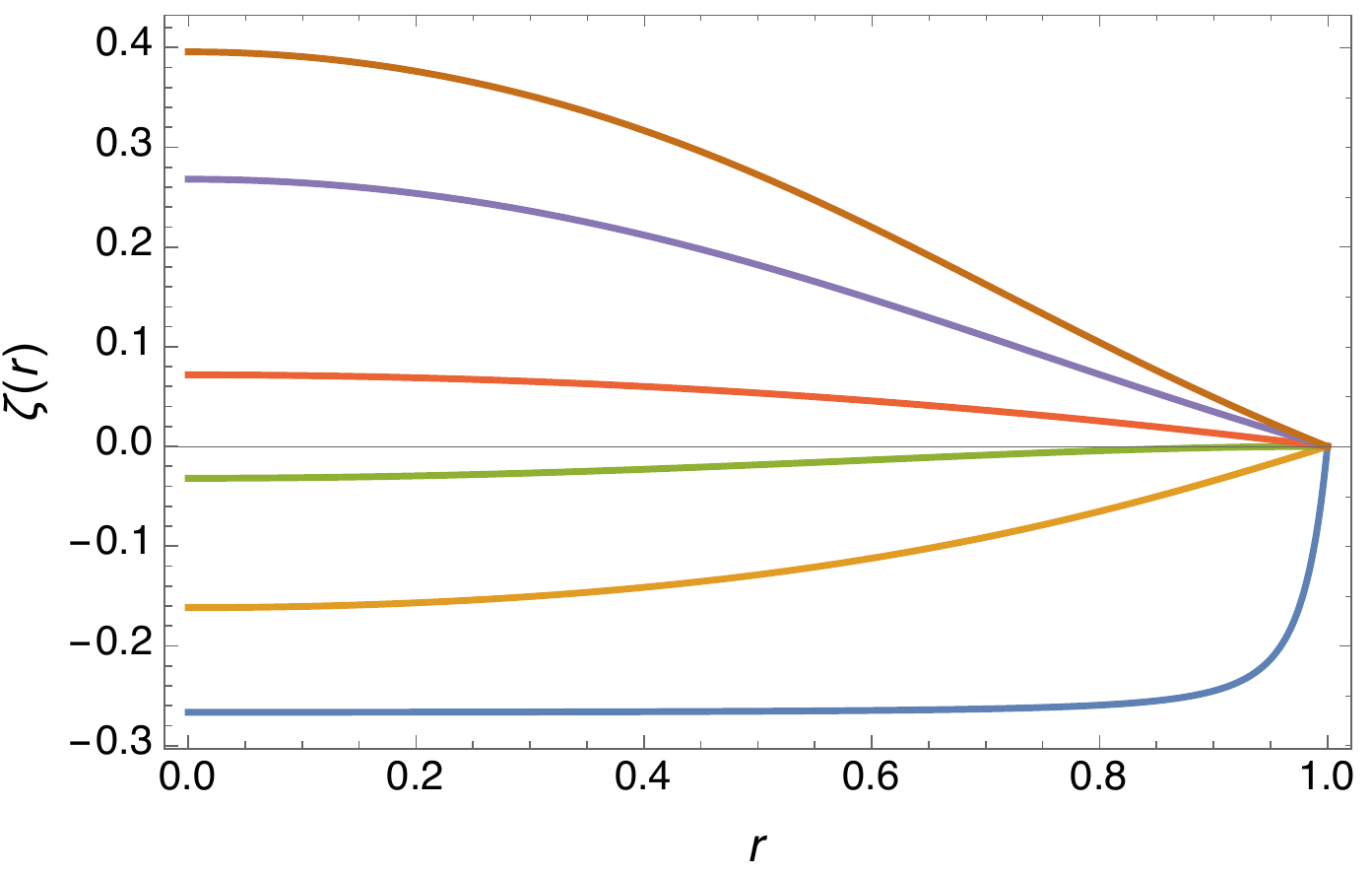}\qquad\qquad \includegraphics[height=1.75 in]{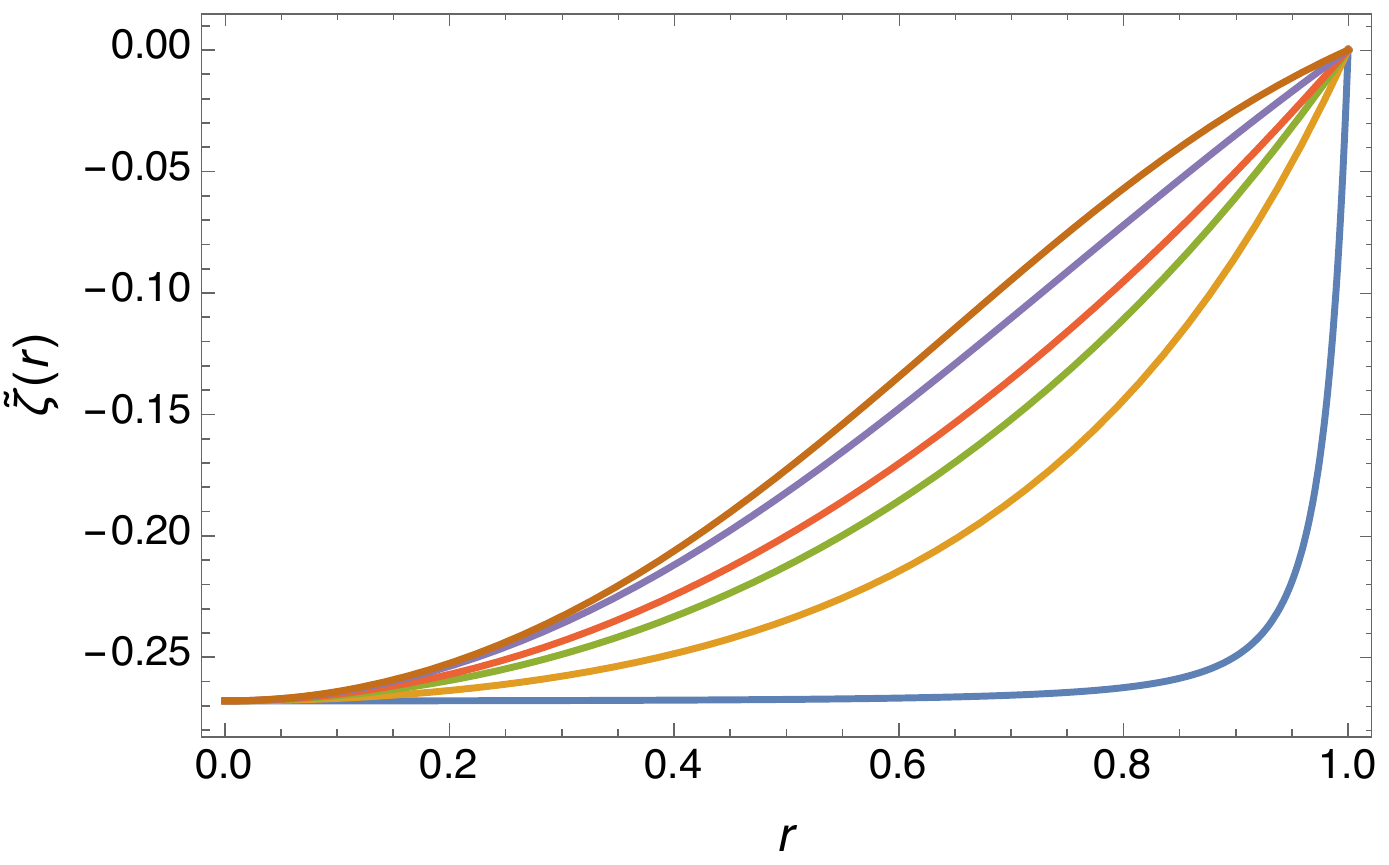}\\[10 pt]
\includegraphics[height=1.75 in]{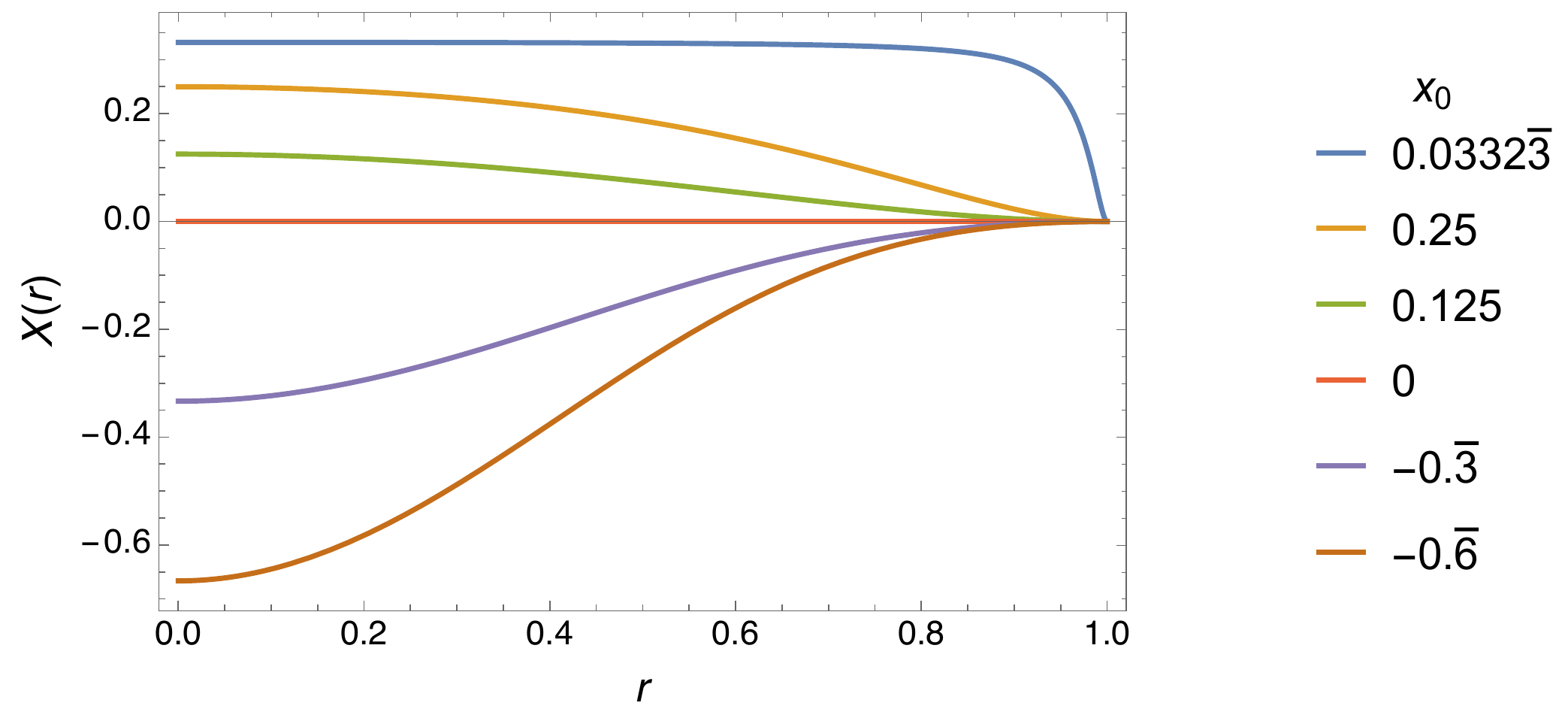}
\caption{A family of solutions of \eqref{zetaeqs} for  different values of $x_0$.  }
\label{fig2}
\end{center}
\end{figure}

\begin{figure}[t]
\centering 
\includegraphics[height=1.75 in]{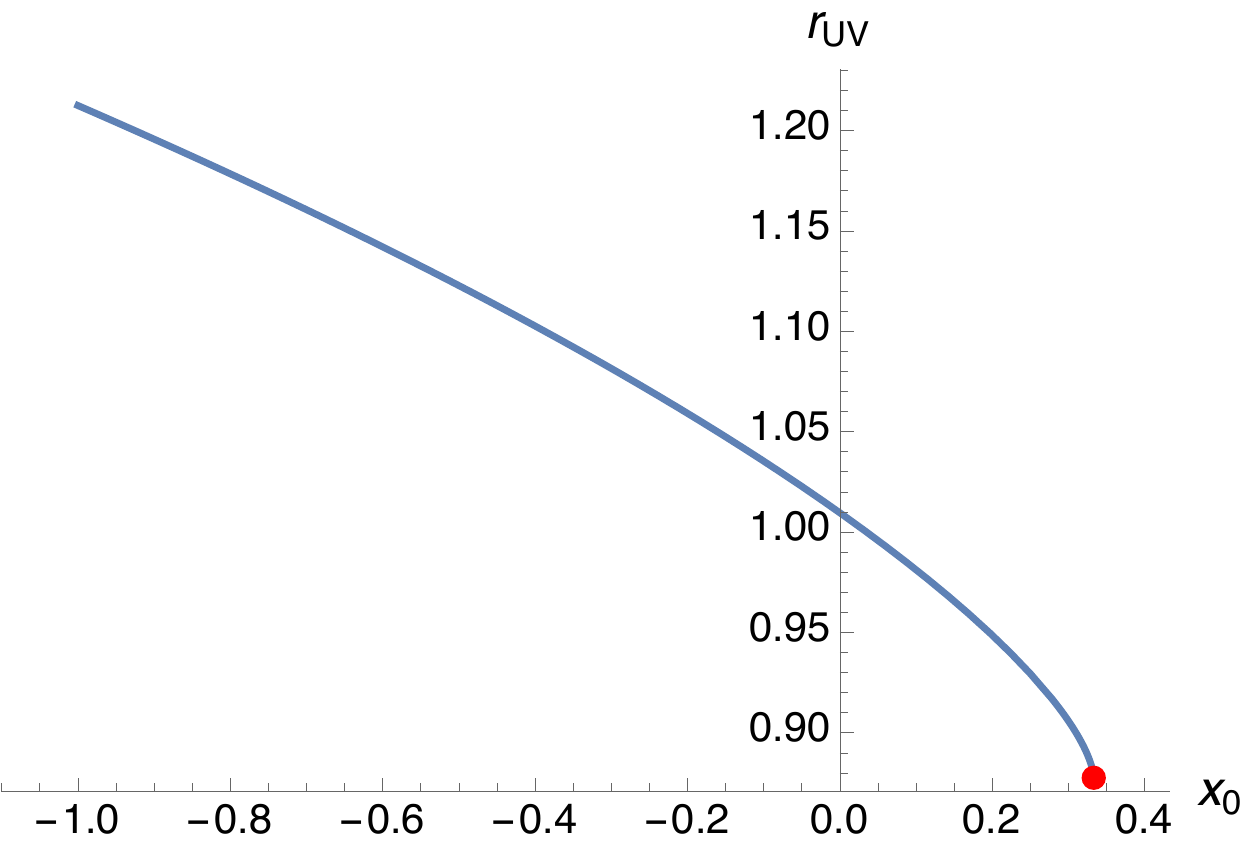}\qquad \includegraphics[height=1.75 in]{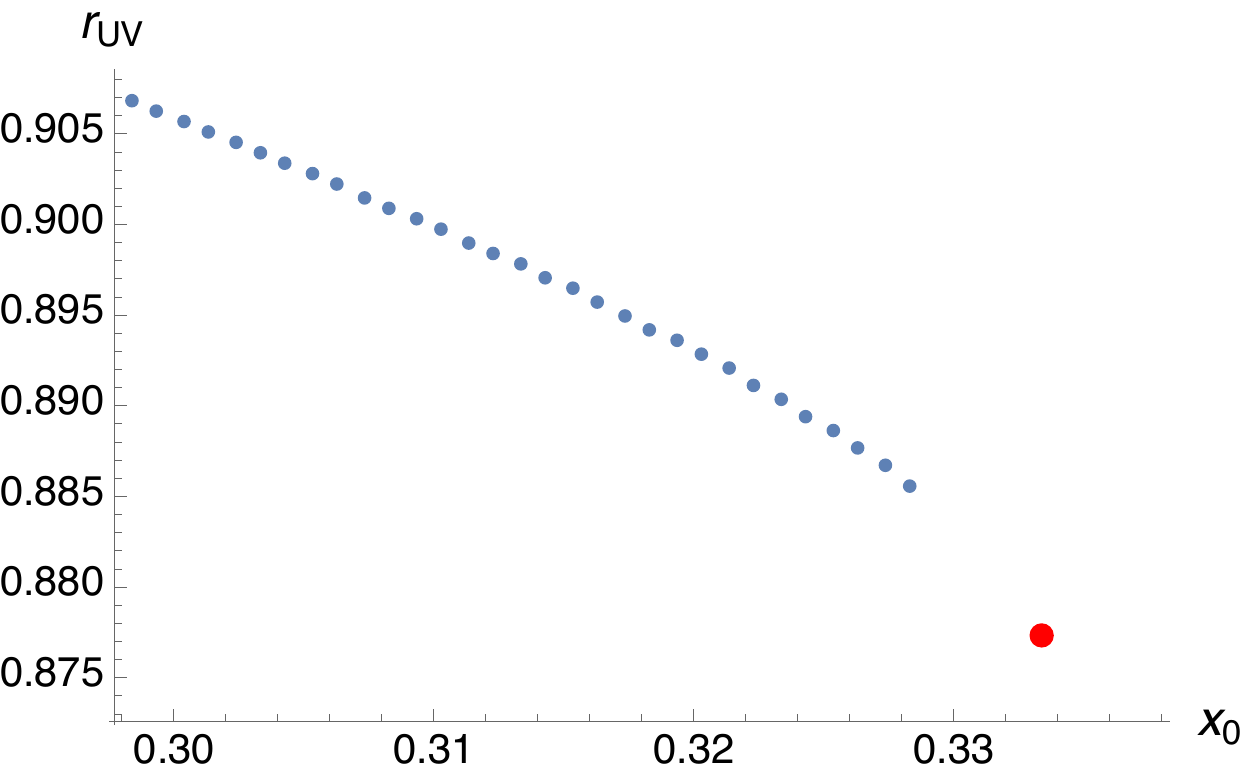}

\caption{Function  $\ruv(x_0)$ obtained by interpolation from numerical solutions. The blue dots on the right are the data points  close to $x_0=1/3$. The red dot denotes $\ruv(1/3)$. \label{fig3} }
\end{figure}

\begin{figure}[t]
\centering 
\includegraphics[height=1.75 in]{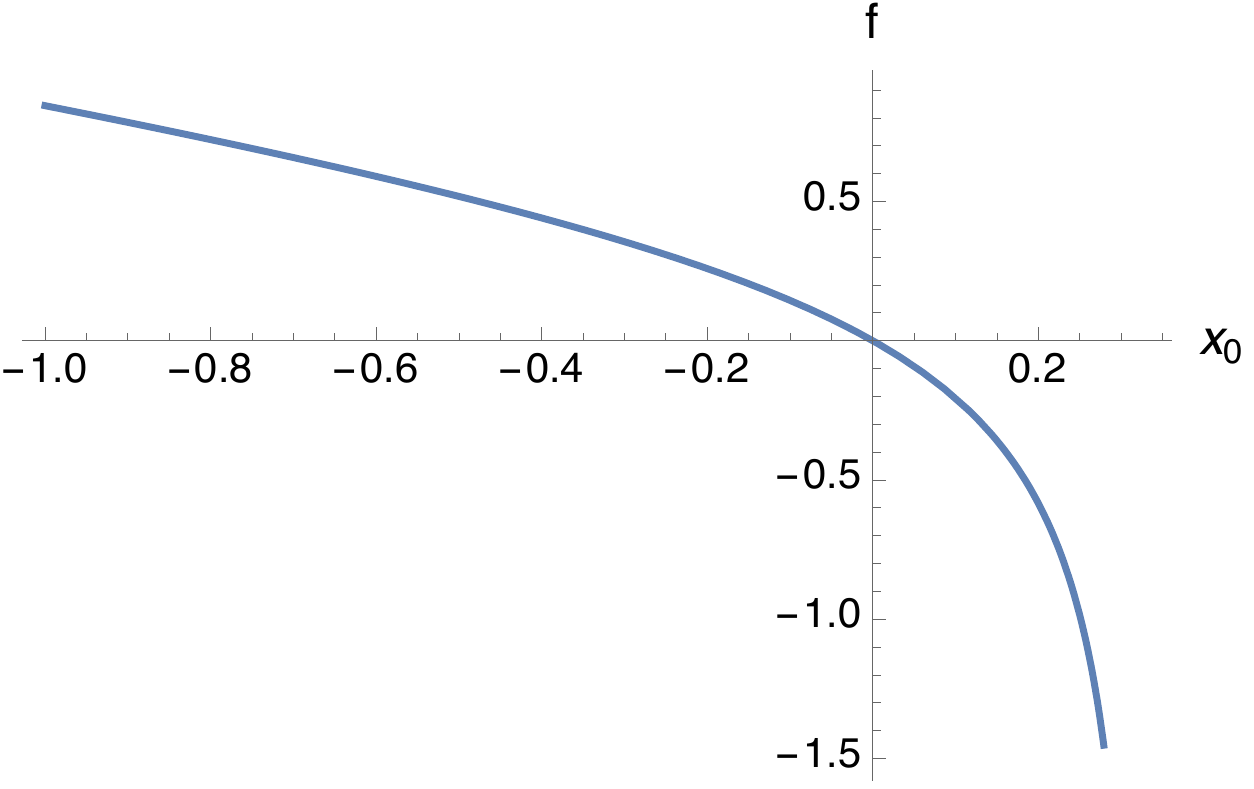}

\caption{Function  $f(x_0)$.\label{fig3a} }
\end{figure}

As expected, by comparing the series expansion of a solution  to a very high order, such as   $\cals O(r^{250})$, with the corresponding solution obtained by a numerical integration, we conclude that the radius of convergence, $R$, of the series solution for all scalars is greater than the UV radius, $\ruv$, for $x_0$ between $-1/3$ and $1/3$. A typical solution in this range is shown in Figure~\ref{fig1}.

For $x_0>1/3$, both the series expansion and/or the numerical integration yield solutions that diverge and do not reach the UV region. 

In Figure~\ref{fig2} we have plotted solutions for the scalars at different values of $x_0$. Those plots suggest that the solution separating the convergent and divergent solutions at $x_0=1/3$ is the  AdS-solution corresponding to the W-point  in Section~\ref{ssec:2ads}. In particular, we see from Figure~\ref{fig3} that the values of $\ruv$ as a function of $x_0$ converge to the AdS value at $x_0=1/3$. The numerics becomes unstable close to $x_0=1/3$ due to the nearly vanishing factor $(\zeta-\bar\zeta)$ in \eqref{zetaeqs}. The data points in the plot on the right were obtained by a series expansion to order $\cals O(r^{300})$.

The value $x_0=-1/3$ is special as it yields another analytic solution, which we describe in Section~\ref{ss:x0m13} below.
 
Finally, for $x_0<-1/3$ we find regular solutions similar to those in Figure~\ref{fig1}, for which $R<\ruv$ and hence one must resort to a numerical integration. 

From the explicit solution we can read-off the relation between the IR parameter, $x_0$, and the UV data given by
\begin{equation}\label{}
a_1\eql a_2\eql a_3\equiv a_s\,, \qquad \tilde a_1\eql \tilde a_2\eql \tilde a_3\equiv \tilde a_s\,,\qquad a\,, \qquad\tilde a\,.
\end{equation}
As expected, see \eqref{aconstr},  we do find that
\begin{equation}\label{}
a_s-\tilde a_s\eql {4\over 3}\qquad \Longrightarrow\qquad \Delta_1\eql \Delta_2\eql \Delta_3\eql {1\over 3}\,.
\end{equation}
However, $a_s$ and $\tilde a_s$ depend on $x_0$,
\begin{equation}\label{}
a_s\eql  {2\over 3}-{2\over 3\sqrt 3}+f(x_0)\,,\qquad \tilde a_s\eql -{2\over 3}-{2\over 3\sqrt 3}+f(x_0)\,,\qquad f(0)\eql 0\,,
\end{equation}
by the same shift, $f(x_0)$,  from their $x_0=0$ values. The function $f(x_0)$ is plotted in Figure~\ref{fig3a}. It diverges at $x_0=1/3$.

\subsubsection{An analytic solution at $x_0=-1/3$.}
\label{ss:x0m13}

This solution is obtained by setting $\zeta\eql -\tilde \zeta$. The  consistency of the first two equations in \eqref{zetaeqs} sets
\begin{equation}\label{}
X\eql -{4\zeta^2\over (1-\zeta^2)^2}\,,\qquad X(0)\eql -{1\over 3}\,.
\end{equation}
Then all three equations in \eqref{zetaeqs} become identical and can be solved by an elementary integration. Using  the normalization \eqref{expAir},  the solution is  given by one of the roots of the following 3rd order polynomial,
\begin{equation}\label{}
\zeta ^3+3 \,\frac{  r^2-r_\text{UV}^2}{r^2+r_\text{UV}^2}\, \zeta ^2-3\, \zeta -\frac{r^2-r_\text{UV}^2}{r^2+r_\text{UV}^2}\eql 0\,,\qquad r_\text{UV}^2\eql {32\over 27}\,.
\end{equation}
It can be written explicitly as
\begin{equation}\label{}
\zeta \eql \cals R-2\,\sqrt{1+\cals R^2}\,\cos\alpha\,,\qquad \alpha\eql {1\over 3}\,\left[\mathop{\rm Arg}(\cals R+i)+\pi\right]\,,\qquad \cals R\eql {r_\text{UV}^2-r^2\over r_\text{UV}^2+r^2}\,.
\end{equation}
In the UV we find
\begin{equation}\label{}
a_s\eql {2\over 3}\,,\qquad \tilde a_s\eql -{2\over 3}\,,
\end{equation}
which shows that 
\begin{equation}\label{}
f(-\hbox{$1\over 3$})\eql {2\over 3\sqrt 3}\,.
\end{equation}
This solution appears to be very special and we were not able to generalize it.

\subsubsection{General numerical solutions}
\label{GNR}

\begin{figure}[t]
\begin{center}
\includegraphics[height=1.85 in]{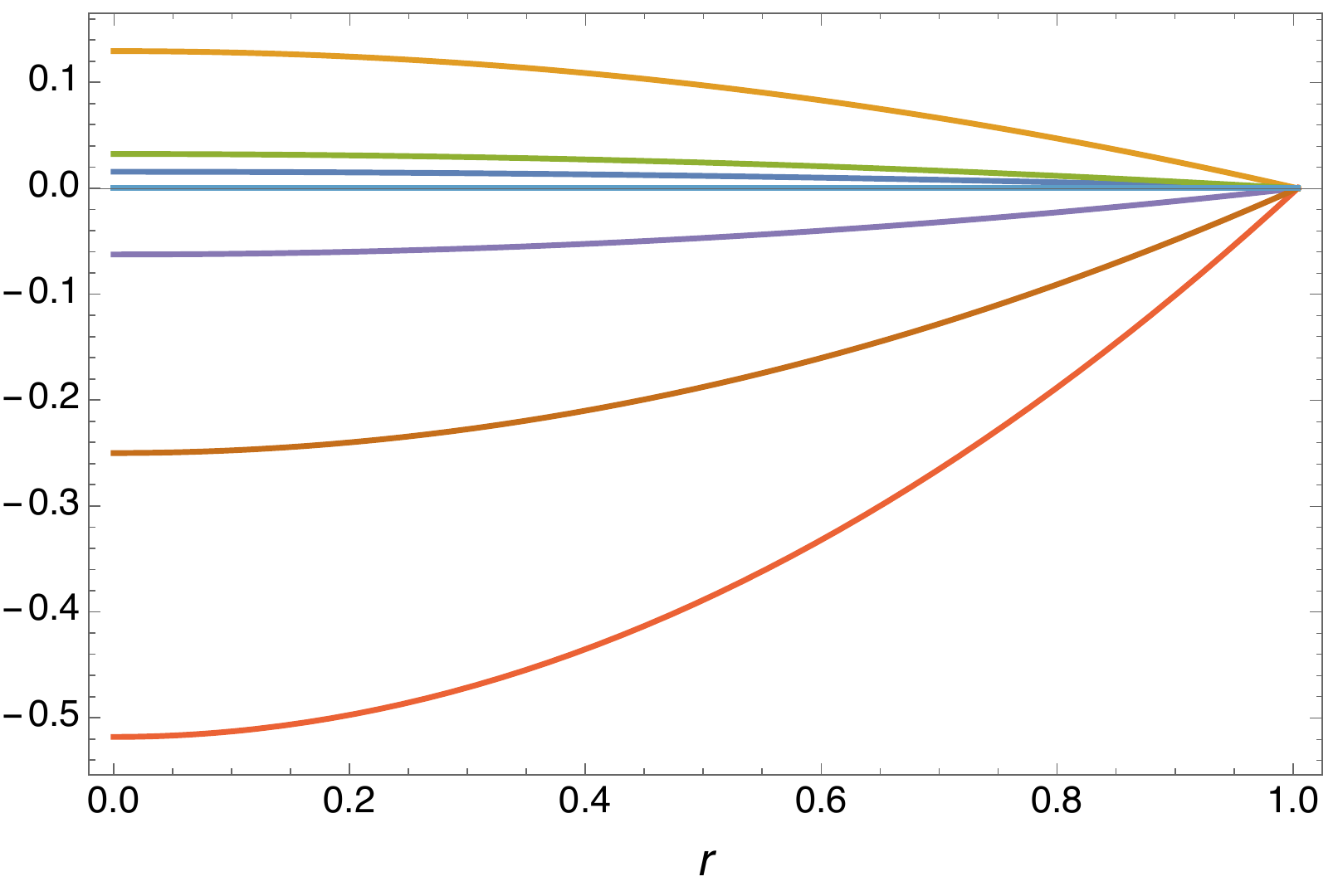}
\includegraphics[height=1.85 in]{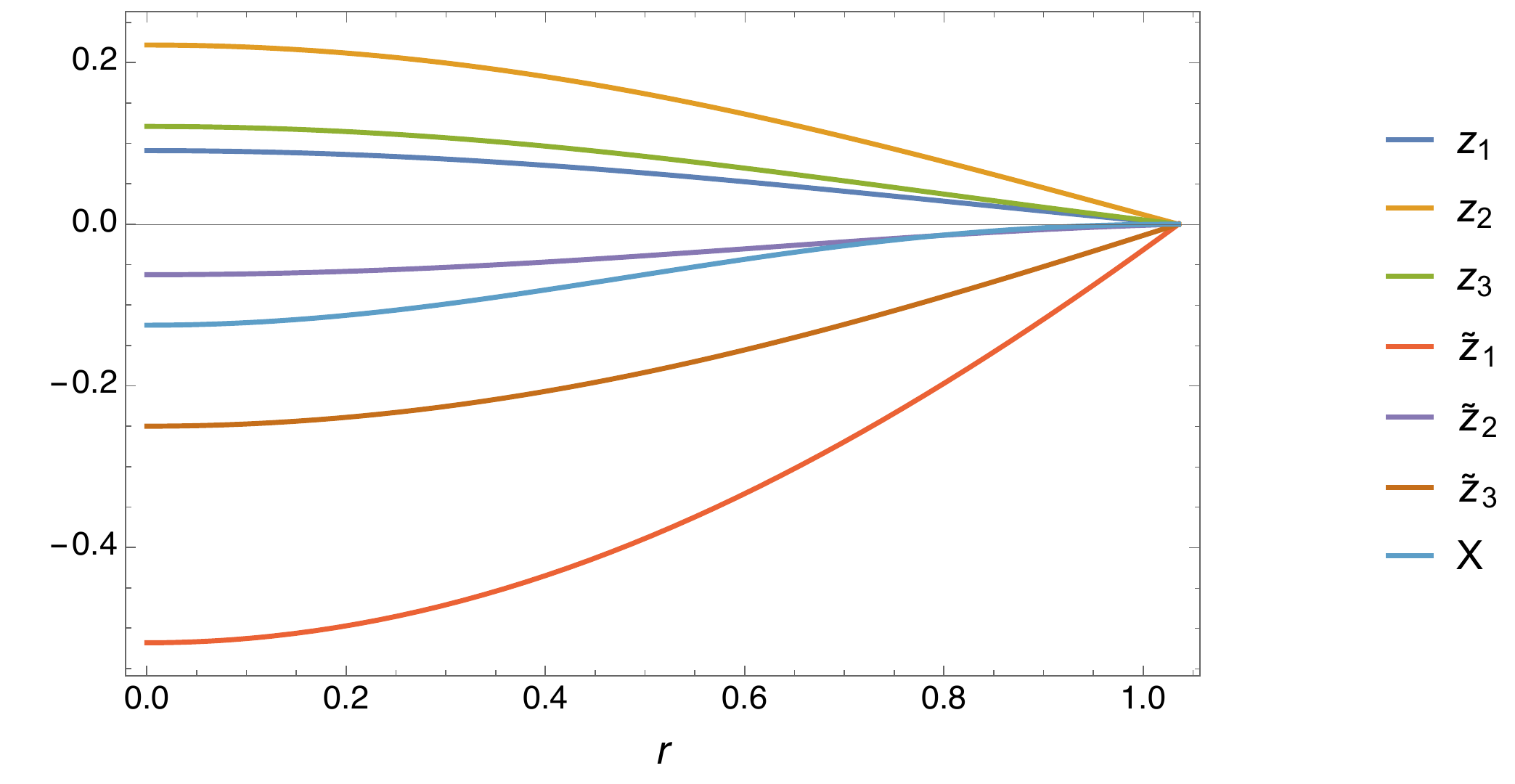}
\caption{Solutions for $x_0=0$ (left) and $x_0=-0.125$ (right) and $\tilde c_1= -0.5181$, $\tilde c_2=-0.0625$, $\tilde c_3=-0.25$.}\label{fig4}\end{center}
\end{figure}

For generic boundary conditions in the IR, the construction of a series solution  proceeds the same way as in the symmetric sector above. We find that starting with random   $\tilde c_i$'s and keeping $|x_0|$ small enough, the power series solution converges as expected. A typical solution is shown in Figure~\ref{fig4}. By combining the power series expansion and numerical routines we were able to explore the relation between the IR and the UV data to establish, based on random sampling, the following results for regular solutions:\footnote{We have restricted our calculations to real $\tilde c_i$ and real $x_0$.}
\medskip

\begin{itemize}\it
\item [(i)] The UV parameters, $\Delta_i$, defined in \eqref{Ddefs} do {\it not} depend on $x_0$ and thus are given by their $x_0=0$ values in \eqref{solDel}.
\item [(ii)] The parameters $ a_i$ and $\tilde a_i$ are shifted from their $x_0=0$ values given in \eqref {fpais} by the {\it same} function $f(\tilde c_i,x_0)$.
\end{itemize}
\medskip

Clearly, the result in (i) follows from (ii). Since there are only two independent $\Delta_i$ and three independent IR parameters, two $\tilde c_i$ and $x_0$, it is not surprising to find a direction in the IR parameter space along which the $\Delta_i$ are constant. However, a priori there is no reason why it should correspond precisely to the IR value, $x_0$, of the hyperscalar. This fact does not appear to be a simple consequence of the symmetry of  equations under a cyclic permutation of the vector scalars as one might have expected.

 Let us denote by $a_i^{(0)}(\tilde c)$ and $\tilde a_i^{(0)}(\tilde c)$ the values of the UV parameters at $x_0=0$ given in \eqref{fpais}. Then
\begin{equation}\label{numobs}
a_i(\tilde c,x_0)\eql a_i^{(0)}(\tilde c)+f(\tilde c,x_0)\,,\qquad \tilde a_i(\tilde c,x_0)\eql \tilde a_i^{(0)}(\tilde c)+f(\tilde c,x_0)\,,
\end{equation}
and we find that the function  $f(\tilde c,x_0)$ depends nontrivially on both $\tilde c_i$ and $x_0$. 

\section{The holographic $S^3$ free energy}\label{sec:HoloRen}

The supergravity solutions constructed above provide the gravitational description of the ABJM theory deformed by real masses, $\delta_i$, in \eqref{deltai} and 
the superpotential mass, $m$,  in \eqref{CPWsuperp}. To test this, in this section we  compute the free energy on the supergravity side and compare with the field theory result  \eqref{FS3CPWgen}. The calculation is somewhat subtle because some of the scalar fields are dual to dimension one operators in ABJM. This means that the usual holographic dictionary, by which  the free energy,  defined as the logarithm of the partition function on $S^3$, is equal to the on-shell supergravity action, is modified. Indeed, the correct relation is that the field theory free energy should be equal to a Legendre transform of the on-shell supergravity action \cite{Klebanov:1999tb,Freedman:2013ryh} (see, also \cite{Freedman:2016yue}).

\subsection{Holographic renormalization and the on-shell action}\label{sec:onshell}

The action in \eqref{bulkS} when evaluated on-shell, in particular on the solutions obtained in Section~\ref{sec:SolBPS}, 
is divergent due to the contributions from the integration over the radial variable, $r$,\footnote{Throughout most of this section, $r$ is a generic radial coordinate and we do not assume any particular gauge for the metric \eqref{Ansatz}.} close to the asymptotically $\mathbb{H}^4$ region. To cancel those divergences one has to add appropriate boundary terms,  or counterterms,  following the standard procedure of holographic renormalization \cite{Skenderis:2002wp}. The radial integral in the action \eqref{bulkS} should then be taken from the IR region at $r=r_\text{IR}$ up to a UV cutoff, $r=r_0$. The boundary terms should similarly be evaluated at $r=r_0$ and only after adding all of these contributions one should remove the cutoff by taking the limit $r_0 \to r_{\rm UV}$.

Implementing this procedure leads to the following regularized action, 
\begin{equation}\label{Sreg}
S_\text{reg} = S_\text{bulk}+S_\text{GH}+S_\mathcal{R}+S_\text{SUSY} \,,
\end{equation}
where $S_\text{bulk}$ is the action \eqref{bulkS} and  
\begin{equation}\label{counterterms}
\begin{split}
S_\text{GH} &\equiv  - \int_{S^3} d\Omega_3 \sqrt{h}\,K \eql - \int_{S^3} d\Omega_3 \,\Big[3 L^3 e^{3A-B} A' \,\Big]_{r=r_0}\,,\\[6 pt]
S_\mathcal{R} &\equiv  \frac{L}{2} \int_{S^3} d\Omega_3 \sqrt{h} \,\mathcal{R} \eql \int_{S^3} d\Omega_3 \,\Big[ 3 L^2 e^A\Big]_{r=r_0}\,, \\[6 pt]
S_\text{SUSY} &\equiv \frac{1}{L} \int_{S^3} d\Omega_3 \sqrt{h}\,\big(\fW\,\widetilde \fW\big)^{1/2} \eql \int_{S^3} d\Omega_3 \, \Big[L^2 e^{3A} \big(\fW\,\widetilde \fW\big)^{1/2}\,\Big]_{r=r_0} \,,
\end{split}
\end{equation}
are the boundary terms evaluated at the cutoff $r_0$. As usual, the prime denotes the derivative with respect to the radial coordinate $r$.

The first two terms in \eqref{counterterms} are the standard Gibbons-Hawking boundary term, $S_\text{GH}$,  and the divergent counterterm, $S_{\cals R}$,  arising from the curvature of the boundary $S^3$-manifold, both evaluated using the boundary metric, $h_{ab}$, induced by the bulk
metric \eqref{Ansatz}.\footnote{Recall that $K$ is the trace of the extrinsic curvature and $\mathcal{R}$ the Ricci scalar of the boundary metric. In the FG-gauge,  $K=\frac{1}{L}\partial_\rho \ln \sqrt{h}$.} 
The last term, $S_\text{SUSY}$, deserves some comments. It depends on  the metric and the scalar fields and contains both divergent terms, quadratic in the scalar fields, as well as finite terms, cubic in the scalar fields, near the $S^3$ boundary. While the  divergent terms can be obtained by the standard holographic renormalization techniques,  the finite terms come with a specific coefficient determined by  a supersymmetric renormalization scheme in the holographic setup. The need for  such finite counterterms in the four-dimensional maximal gauged supergravity and its truncations  was emphasized in \cite{Freedman:2013ryh} and \cite{Freedman:2016yue}. The  $S_\text{SUSY}$ counterterm in \eqref{counterterms} is an obvious generalization of the corresponding counterterm in the STU-model in \cite{Freedman:2013ryh}. It also follows from the general result in Appendix C of \cite{Freedman:2016yue} applied to the present truncation.

By converting the boundary terms to a total derivative with respect to $r$ and integrating over the sphere, we can rewrite the regularized action \eqref{Sreg} as 
\begin{equation}\begin{split}\label{Sregexpl}
S_\text{reg} & =\int^{r_0}_{r_\text{IR}} dr\int_{S^3} d\Omega_3\,\mathcal{L}_{\text{reg}} \\ 
& =\textrm{vol}_{S^3} L^3 \int^{r_0}_{r_\text{IR}} dr\Bigg[\,  e^{3A-B} \bigg(  -3 (A')^2-{3 \over L^{2}}\, e^{2(B-A)} \\ &\hspace{139 pt}+  \sum_{i=1}^3 \frac{z_i'\zt_i'}{(1-z_i \zt_i)^2} + \frac{z' \zt'}{(1-z \zt)^2}  
+ \frac{e^{2B}}{2L^2}\, \mathcal{P}\bigg)\\ & \hspace{195 pt}+ {1\over L}\,\left(3 e^A+e^{3A}\big(\fW\,\widetilde \fW\big)^{1/2} \right)' \, \Bigg] \,,
\end{split}\end{equation}
where $\textrm{vol}_{S^3}=2\pi^2$ is the volume of the unit three-sphere and we have introduced the regularized Lagrangian $\mathcal{L}_{\text{reg}}$. For  regular solutions in Section~\ref{sec:SolBPS} that are of interest here, the metric function  $e^A$ vanishes at $r=r_\text{IR}$\footnote{See, e.g., \eqref{Aat0} with $r_\text{IR}=0$.} while the scalar fields are finite. This ensures that by recasting the counterterms \eqref{counterterms} into bulk integrals we do not introduce any additional terms in the IR.

Since we are interested in evaluating the regularized action \eqref{Sregexpl} on-shell, we can now employ the BPS equations and the equations of motion to further simplify its form. To this end, we note that the BPS equation \eqref{AsqEqs} and the equation of motion for $A$ in \eqref{EOMs} can be used to rewrite $\fW\,\fWt$ and the kinetic terms for the scalar fields in terms of the metric functions $A$ and $B$.
Furthermore,  the BPS equations imply the identity
\begin{equation}
\frac{z_i'\zt_i'}{(1-z_i \zt_i)^2} + \frac{z' \zt'}{(1-z \zt)^2} = \frac{e^{2B}}{2L^2}\left(\mathcal{P}+\frac{3}{2}\fW \fWt\right)\,,
\end{equation}
which allows us to rewrite also the potential in terms of the metric functions.  Finally, collecting the resulting terms into a total derivative and then rewriting the latter as a new boundary term, we are left with the following result for the on-shell action,
\begin{equation}\label{Sonshell}
\begin{split}
S_\text{on-shell} & \equiv \lim_{r_0\to \ruv} S_\text{reg}\\
& \eql \textrm{vol}_{S^3}L^2 \,\lim_{r_0\to \ruv}\Bigg[ \int_{r_\text{IR}}^{r_0} dr \Big [-\frac{2}{L}e^{A+B}\Big]\\
& \hspace{100 pt}+\Big[3 e^{A} -2L e^{3A-B}A' +2e^{3A}\sqrt{L^2 e^{-2B}(A')^2-e^{-2A}}\,\Big]_{r=r_0}\Bigg]\,,
\end{split}
\end{equation}
which involves the metric functions only. Note that \eqref{Sonshell} is manifestly invariant under reparametrizations of the radial coordinate and is thus valid in any gauge. 

Although the boundary contribution in the last line in \eqref{Sonshell} may seem somewhat involved, one has to remember that only the singular and the finite terms  as $r_0\to \ruv$ contribute to the on-shell action. In turn, those terms are determined by the UV asymptotics of a solution leading to a rather simple result. To illustrate this, consider \eqref{Sonshell} in the conformal gauge \eqref{conf}. Using \eqref{ruv3}, the UV expansion of the boundary term is 
\begin{equation}\label{Bryexp}
3 e^{A} -2r e^{2A}A' +2e^{2A}\sqrt{r^2(A')^2-1}\eql -{2\ruv\over r-\ruv}-1+\cals O(r-\ruv)\,.
\end{equation}
It is rather remarkable that while the individual terms on the left hand side depend on the regular terms in  the expansion \eqref{ruv3}, the contribution from those regular terms cancels out so that only the singular part of the UV expansion   \eqref{ruv3} is needed to obtain \eqref{Bryexp}. Then the  on-shell action \eqref{Sonshell} in the conformal gauge is simply given by 
\begin{equation}\label{}
S_\text{on-shell} \eql  -2\,\textrm{vol}_{S^3}L^2 \Bigg[\int_0^{\ruv} dr\,\left({e^{2A}\over r}-{\ruv\over (r-\ruv)^2}\right)-{1\over 2}\,\Bigg]\,,
\end{equation}
where the integral is now convergent.

It is now straightforward to compute the on-shell action for the solutions found in Section~\ref{sec:SolBPS}. In particular, for the analytic solution   in the limit of the vanishing hyperscalars \eqref{FPsol}-\eqref{expAan} we find
\begin{equation}\label{Sonshx00}
S_\text{on-shell}(\tilde c_i, x_0=0)\eql 2\,\textrm{vol}_{S^3}L^2 \,{1+\tilde c_1\tilde c_2\tilde c_3\over 1-\tilde c_1\tilde c_2\tilde c_3}\,,
\end{equation}
which agrees with \cite{Freedman:2013ryh}. For generic solutions that are parametrized by the constants $\tilde c_i$ and $x_0=c\,\tilde c$, the on-shell action can be evaluated only numerically and we find that it depends nontrivially on all  parameters, in particular on $x_0$.   

\subsection{Canonical conjugates and the Legendre transform} 
\label{sec:conjugate}

The scalar fields of the $\cals N=8$ gauged supergravity in four dimensions are 35~scalars and 35~pseudoscalars that are respectively dual to scalar bilinear operators of conformal dimension one and to fermionic bilinear operators of dimension two in the ABJM theory. As was first emphasized in \cite{Klebanov:1999tb}, following the earlier work on quantization in AdS backgrounds  \cite{BREITENLOHNER1982249}, this leads to different holographic dictionaries  for the coefficients in the UV expansions for the scalars and the pseudoscalars. For the pseudoscalars,  the leading term in \eqref{FGexpand}  is as usual the ``source'' and the subleading  term is the ``vev'' for the dual operator, but the roles are ``altered'' when the dual operator has dimension one. More precisely, one must Legendre transform the on-shell action with respect to the scalars upon which the leading term of the conjugate field becomes the ``source'' and the subleading term the ``vev.''  Clearly, the same rules apply to any truncation of the maximal theory.

Tracing back the $\mathcal{N}=8$ supergravity origin of the fields in our consistent truncation, one finds that the linear combinations $z_i-\zt_i$ as well as both   $z$ and $\tilde{z}$ are pseudoscalars, while the linear combinations $z_i+\zt_i$ correspond to scalars.\footnote{See, Appendix B in \cite{Bobev:2018uxk}, in particular equations (B.6)-(B.8) with $\zeta_1=0$ and $\zeta_2=z$. One can also deduce the parity of the scalar fields by rewriting the supersymmetry variations \eqref{varr32}, \eqref{var12} and \eqref{var12b} in terms of Majorana spinors.}  This means that, similarly as in \cite{Freedman:2013ryh}, we must perform the Legendre transform with respect to the combination $z_i+\zt_i$ of the scalar fields.

To implement the Legendre transform, we start by computing the canonical conjugates for the scalar fields. For a more direct comparison with similar calculations in  \cite{Freedman:2013ryh,Freedman:2016yue}, we work in 
the  Fefferman-Graham gauge \eqref{FG} for which the UV expansion of the scalar fields is given in \eqref{FGexpand}.
The canonical conjugate of the leading term, $a_i$, is then defined as
\begin{equation}\label{conja}
\mathfrak{a}_i \equiv - \lim_{\rho_0\to\infty}\frac{\delta S_\text{reg}[a_i,\tilde a_i,a,\tilde a]}{\delta a_i} = - \lim_{\rho_0 \rightarrow \infty} e^{-\rho_0} \Pi_{z_i}(\rho_0) \,.
\end{equation}
Here $\rho_0$ is the UV cut-off   and 
\begin{equation}\label{momentum} 
\Pi_{z_i}(\rho) =\frac{\partial \mathcal{L}_\text{reg}}{\partial(\partial_\rho z_i)} \,,
\end{equation}
where $\mathcal{L}_\text{reg}$ is defined in \eqref{Sregexpl}.
The canonical conjugates of $\at_i$, $a$ and $\at$ are defined in a similar way.
A direct calculation gives the following result\footnote{There is a relative minus sign between our results and those of \cite{Freedman:2013ryh}. We expect that  a minus sign is missing in (6.12) of \cite{Freedman:2013ryh}, as it is also necessary to obtain (6.21) from (6.19) and (6.20) in \cite{Freedman:2013ryh}.}
\begin{equation}\begin{split}\label{canonconj}
\ma_i &=\frac{L^2}{8}\left( \bt_i+\frac{a_1 a_2 a_3}{a_i}-\frac{a \at}{2}\right) \,, \\
\mat_i &=\frac{L^2}{8}\left( b_i+\frac{\at_1 \at_2 \at_3}{\at_i}-\frac{a \at}{2}\right) \,, 
\end{split}\end{equation}
and
\begin{equation}\begin{split}\label{canonconjaat}
\ma &=\frac{L^2}{8}\left(\bt-\frac{\at}{2}\left(a_1+a_2+a_3+\at_1+\at_2+\at_3\right)\right) \,,\\
\mat &=\frac{L^2}{8}\left(b-\frac{a}{2}\left(a_1+a_2+a_3+\at_1+\at_2+\at_3\right)\right) \,.
\end{split}\end{equation}
As expected,  modulo the quadratic  terms on the right hand sides  in \eqref{canonconj}-\eqref{canonconjaat}, the canonical conjugate of the leading term in \eqref{UVexp} is   the subleading term in \eqref{UVexp}, which is an explicit realization of the exchange between a ``source'' and a ``vev.'' However, see also \cite{Freedman:2016yue}, the exchange is not exact; there are corrections due to the quadratic terms that arise from the counterterm, $S_\text{SUSY}$. In this sense, the  precise form of the canonical conjugates in \eqref{canonconj}-\eqref{canonconjaat} is determined by supersymmetry.

The Legendre transformed on-shell action is now given by 
\begin{equation}\label{Jinit}
J_\text{on-shell} = S_\text{on-shell}+\frac{1}{2}\int_{S^3} d\Omega_3 \sum_{i=1}^3 (a_i+\at_i)(\ma_i+\mat_i) \,,
\end{equation}
and we need to  evaluate it explicitly on regular solutions of the BPS equations  \eqref{zzjEqs}-\eqref{AsqEqs}. 

The UV analysis of the BPS equations in Section~\ref{sec:UV} established explicit relations  between the leading and subleading terms in \eqref{UVexp}, which we can now use to simplify \eqref{canonconj}-\eqref{Jinit}.  Indeed, using \eqref{aconstr}-\eqref{abrels2} in \eqref{canonconj} , we find that
\begin{equation}\label{vanfa}
\ma_i = - \frac{L^2}{4} \at_i \,,\qquad  \mat_i  = \frac{L^2}{4} a_i \,.
\end{equation}
In particular,
\begin{equation}\label{sor12}
\ma_i\pm \mat_i\eql {L^2\over 4}(a_i\mp \tilde a_i)\,,
\end{equation}
which reflects the fact that supersymmetry fixes the relative coefficients between the sources of dimension one and dimension two operators in the deformation of the ABJM Lagrangian. 

Using \eqref{aconstr}-\eqref{abrels2} in \eqref{canonconjaat}, one finds that  
\begin{equation}\label{vanaat}
\ma\eql \mat \eql 0\,.
\end{equation}
This suggestive result has an interpretation in the dual field theory. The scalar fields $z$ and $\tilde{z}$ are dual to the fermionic bilinear operators sourced by the superpotential deformation in \eqref{CPWsuperp}. Thus their ``sources'' in the UV expansion \eqref{FGexpand} are proportional to the superpotential mass parameter, $m$. From the results discussed in Section~\ref{sec:CFT}, it is clear that the free energy as well as other supersymmetric observables, which can be computed by supersymmetric localization of the path integral, do not depend continuously on the parameter, $m$. This can be understood as a Ward identity for correlation functions and the vanishing of the canonical conjugates in  \eqref {vanaat} can be viewed as the supergravity counterpart of this Ward identity.

\subsection{Free energy}\label{sec:FS3SUGRA}

The Legendre transformed on-shell action, $J_\text{on-shell}$, is a function of $\ma_i+\mat_i$, $a_i-\at_i$, $a$ and $\tilde a$. Using \eqref{sor12}, we can rewrite \eqref{Jinit} as
\begin{equation}\label{J}
J_\text{on-shell} = S_\text{on-shell} + \textrm{vol}_{S^3} \frac{L^2}{8} \sum_{i=1}^3 (a_i^2-\at_i^2) \,,
\end{equation}
so that it becomes a function of the UV parameters, a priori, $a_i$, $\tilde a_i$, $a$ and $\tilde a$. The problem that we now face is to evaluate \eqref{J} on the space of regular solutions to the BPS equations. Those correspond to 
a subspace of allowed UV parameters which we explored numerically in Section~\ref{GNR} with the main results summarized in   (i) and (ii).

Since, unlike in \cite{Freedman:2013ryh}, we do not know solutions to our BPS equations in a closed form, we cannot evaluate \eqref{J} directly. It is then natural to 
first try to determine how $J_\text{on-shell}$ varies within the space of solutions.  This is a much easier problem because  the variation of the on-shell action is effectively a boundary calculation.

Indeed, by repeating the same steps as in Appendix C.3 in \cite{Bobev:2013cja} and using \eqref{conja} and \eqref{momentum} along the way, we obtain the following general result for the variation of the regularized  action,
\begin{equation}\begin{split}\label{varSreg}
\frac{d S_\text{reg}}{d\mu} &= -\int_{S^3} d\Omega_3 \left[\,\sum_{i=1}^3\Big(\ma_i \frac{\partial a_i}{\partial\mu}+\mat_i \frac{\partial \at_i}{\partial\mu}\Big)+ \ma \frac{\partial a}{\partial\mu}+\mat \frac{\partial \at}{\partial\mu} \,\right]\,,
\end{split}\end{equation}
where $\mu$ parametrizes any variation of the UV data. In particular, we can apply  \eqref{varSreg}
to variations of the on-shell action along a family of regular solutions parametrized by $\mu$. 
Including the Legendre transform and using \eqref{vanfa}, we then find
\begin{equation}
\begin{split}\label{dJdmu}
\frac{d J_\text{on-shell}}{d\mu} &= \frac{L^2}{4} \int_{S^3} d\Omega_3 \sum_{i=1}^3\left[\,\Big(\at_i \frac{\partial a_i}{\partial\mu}-a_i \frac{\partial \at_i}{\partial\mu}\Big) + \frac{1}{2} \frac{\partial}{\partial \mu} \big[(a_i+\at_i)(a_i-\at_i)\big]\right] \\[6 pt]
&= \textrm{vol}_{S^3} \frac{L^2}{4}\sum_{i=1}^3(a_i+\at_i)\frac{\partial}{\partial \mu}(a_i-\at_i)\,.
\end{split}
\end{equation}  
Next, recall that $a_i-\tilde a_i$ are proportional to  $\Delta_i$ defined in \eqref{Ddefs} and that, by our numerical analysis in Section~\ref{GNR}, the latter do  not depend on $x_0=c\,\tilde c$, where $c$ and $\tilde c$ are the values of the hyperscalars in the IR. This means that, in fact, 
\begin{equation}\label{Jindx0}
\frac{d J_\text{on-shell}}{d x_0}\eql 0\,,
\end{equation}
and hence $J_\text{on-shell}$ is determined, at least locally, by its $x_0=0$ value, for which it can be calculated exactly given \eqref{Sonshx00} and \eqref{fpais}. The result is 
\begin{equation}\label{Jincs}
J_\text{on-shell}\eql  2 \,  \textrm{vol}_{S^3} L^2\,\frac{(1-\ct_1^2)(1-\ct_2^2)(1-\ct_3^2)}{(1-\ct_1\ct_2\ct_3)^2} \,,
\end{equation}
which is the same as (6.21) in  \cite{Freedman:2013ryh} except that the parameters $\tilde c_i$ are now subject to the cubic constraint \eqref{cubic}. Using \eqref{solDel}, we can rewrite \eqref{Jincs} as
\begin{equation}\label{Jnum}
J_\text{on-shell}\eql 2\,\textrm{vol}_{S^3} L^2\sqrt{\Delta_1\Delta_2\Delta_3} \,,
\end{equation}
which is the main result of our supergravity calculation.

We have also confirmed \eqref{Jnum} by extensive numerical checks in which we evaluated $J_\text{on-shell}$ given by \eqref {J} directly on numerical solutions. In particular, we find that a nontrivial dependence of the on-shell action on the hyperscalar parameter, $x_0$, is always cancelled by the corresponding contribution from the additional terms in \eqref {J} due to the Legendre transform. Combined with the observations in Section~\ref{GNR}, this implies that the dependence of $S_\text{on-shell}$ on $x_0$ is given by the  function $f(\tilde c_i,x_0)$ in \eqref{numobs},

The comparison with the field theory partition function is now straightforward. In field theory, the real mass parameters, $\delta_i$, in \eqref{deltai} are couplings to dimension two operators \cite{Freedman:2013ryh} which, on the supergravity side, are sourced  by $a_i-\at_i$.  We thus expect the relation $a_i - \at_i = n \delta_i$ for some constant $n$. Since the supergravity constraint \eqref{aconstr} becomes identical to the field theory constraint \eqref{deltaconstr} only if we set $n=8$, this leads  to the map
\begin{equation}\label{adelta}
a_i - \at_i = 8 \,\delta_i \,,
\end{equation}
or, equivalently, after we use  \eqref{Deltadelta} and \eqref{Ddefs}, to
\begin{equation}\label{Deltarel}
\Delta_1 = \Delta_{A_2}\,, \qquad \Delta_2 = \Delta_{B_1}\,, \qquad \Delta_3 = \Delta_{B_2}\,.
\end{equation}
This also makes   the relation between the field theory constraint in \eqref{DeltaconstrmABJM} and the constraint \eqref{Delid} in supergravity manifest.

To complete the comparison between the field theory and supergravity results, we need the relation between the AdS scale, $L$, and rank of the gauge group $N$,\footnote{See, e.g.,  \cite{Bobev:2018uxk} for a discussion of this relation in the conventions of this paper.}
\begin{equation}\label{GN}
\textrm{vol}_{S^3} L^2=\frac{\pi}{3\sqrt{2}} N^{3/2} \,.
\end{equation}
Then  the holographic free energy \eqref{Jnum} can be written as
\begin{equation}\label{JCPWgen}
J_\text{on-shell} =\frac{4\sqrt{2}\pi}{3}N^{3/2}\sqrt{\Delta_{1}\Delta_{2}\Delta_{3}} \;,
\end{equation}
and, given \eqref{Deltarel}, is manifestly equal to the field theory free energy \eqref{FS3CPWgen} obtained via supersymmetric localization.

\section{Conclusions}
\label{sec:Conclusions}

In this paper we presented a non-trivial precision test of the gauge/gravity duality by successfully comparing the large $N$ result for the free energy of the ABJM theory deformed by real and superpotential masses with the corresponding calculation in supergravity. Our supergravity construction extends and generalizes the results of \cite{Freedman:2013ryh}. We  turn on two additional scalar fields in the bulk that are dual to the superpotential mass and find novel holographic RG flow solutions in Euclidean signature. After carefully applying holographic renormalization and alternative quantization,  the on-shell action of these supergravity solutions agrees with the results for the free energy of the theory computed by supersymmetric localization. There are several interesting avenues for extending these results.

The deformation of ABJM theory with a single superpotential mass term, \eqref{CPWsuperp},  can be generalized by turning on superpotential masses for all four chiral superfields. This general superpotential deformation preserves $\mathcal{N}=2$ supersymmetry and can be combined with a deformation by non-vanishing real mass terms. The free energy of such theory on $S^3$ can be computed  by supersymmetric localization and the result is analogous to the one in \eqref{DeltaconstrmABJM} and \eqref{FS3CPWgen}; the real mass parameter for a chiral superfield with a superpotential mass term is fixed as in \eqref{DeltaconstrmABJM} and the free energy as a function of the remaining undetermined real masses is obtained from \eqref{FS3}. Although it should be possible to reproduce this simple QFT result from supergravity by following the approach presented in this paper,   the details will most likely  be technically challenging. The reason is that for each new superpotential mass term in the ABJM theory, one has to enlarge the supergravity truncation of the four-dimensional $\mathcal{N}=8$ supergravity to include appropriate scalar fields in the bulk. Then one has to construct Euclidean supergravity solutions within the larger truncation, perform holographic renormalization, and compute the  on-shell action. It would be quite interesting to understand whether one could somehow circumvent this brute force supergravity calculation to arrive at the final simple QFT result in a more direct manner.

We have been somewhat conservative in choosing the range of parameters for our supergravity solutions by insisting that the metric be  real and the scalar fields   lie inside the unit disk. Perhaps there are more general complex saddle points of the Euclidean supergravity action that have a physical interpretation. In the dual QFT this amounts to a careful analysis of the range of the parameters $\delta_i$ in \eqref{deltai} on the complex plane for which there is a saddle point of the localization matrix model such that the free energy scales as $N^{3/2}$. It would be interesting to understand this  better both in supergravity and in field theory. 

One can view our results, as well as those in \cite{Freedman:2013ryh}, as a first step to harness the power of localization in the context of non-conformal holography for the ABJM theory on $S^3$. A more ambitious goal is to extend this construction beyond the leading order in the large $N$ approximation. This is challenging both on the field theory and on the gravity side. In \cite{Nosaka:2015iiw}, the partition function of the ABJM theory with two non-vanishing real masses was computed to all orders in the $1/N$ expansion, but a general result for arbitrary real and superpotential masses is not currently available. On the gravity side the problem can perhaps be addressed by first uplifting the four-dimensional solutions constructed above, as well as the ones in \cite{Freedman:2013ryh}, to eleven dimensions and then understanding modifications by higher curvature corrections. While the uplift to eleven dimensions is in principle algorithmic,\footnote{See \cite{Bobev:2018hbq} for the explicit uplift to type IIB supergravity of a similar five-dimensional solution with an $S^4$ boundary found in \cite{Bobev:2013cja}.} taking into account higher curvature corrections is presumably a tall order. 
Perhaps a more accessible problem is to perform a one-loop analysis of the eleven-dimensional bulk solution, along the lines of \cite{Bhattacharyya:2012ye}, and in this way compute the logarithmic term in the $1/N$ expansion from supergravity and then compare to the supersymmetric localization result.

Finally, we would like to note that a supergravity truncation with the same scalar fields content, but a different gauging from the one studied here, arises from a compactification of the massive IIA supergravity on $S^6$ \cite{Guarino:2015vca}. This truncation was studied further in \cite{Kim:2018sdw} with the goal of constructing Euclidean $S^3$ solutions of the type discussed above. Such solutions should have an analogous interpretation to the ones presented here as well as in \cite{Freedman:2013ryh}. Namely, they should be holographically dual to deformations of  the three-dimensional $\mathcal{N}=2$ SCFT studied in \cite{Guarino:2015jca} by real and superpotential mass terms. It should be possible to construct these supergravity solutions explicitly and compare their on-shell action to the localization result for the free energy as a function of the real masses presented in \cite{Guarino:2015jca}, see also Appendix A of \cite{Azzurli:2017kxo}.

\bigskip
\bigskip
\leftline{\bf Acknowledgements}
\smallskip
\noindent We are grateful to Daniel Z. Freedman, Fri\dh rik Freyr Gautason, Silviu S. Pufu, and Alberto Zaffaroni for interesting discussions. The work of NB is supported in part by an Odysseus grant G0F9516N from the FWO. The work of VSM is supported by a doctoral fellowship from the Fund for Scientific Research - Flanders (FWO) and in part by the European Research Council grant no. ERC-2013-CoG 616732 HoloQosmos. NB and VSM are also supported by the KU Leuven C1 grant ZKD1118 C16/16/005. KP and FR are supported in part by DOE grant DE-SC0011687.

\appendix
\section{The U(1)$^2$-invariant truncation}

\label{appendixA}

In this appendix we summarize the results in \cite{Bobev:2018uxk} for the action and supersymmetry variations for    the $\rm U(1)^2$-invariant  truncation of the $\cals N=8$, $d=4$ gauged supergravity \cite{deWit:1982bul}.  A more detailed discussion of that truncation can be found in Section 3 and Appendix B in \cite{Bobev:2018uxk} and the references therein.

\subsection{The action}

As shown in \cite{Bobev:2018uxk}, the $\rm U(1)^2$-invariant truncation of the maximal gauged supergravity in four dimensions  is the $\cals N=2$ gauged supergravity coupled to the three Abelian vector multiplets of the STU-model \cite{Duff:1999gh,Cvetic:1999xp,Behrndt:1996hu} and the universal hypermultiplet (see, e.g., \cite{BrittoPacumio:1999sn} and the references therein). One can also impose an additional $\rm U(1)$-symmetry on the bosonic fields,  which projects out one of the scalars in the hypermultiplet. Finally, the four vector fields of the $\rm U(1)^4$ gauge symmetry can be consistently  set to zero.\footnote{See, the comment after \eqref{Kgflds} below.} Then the remaining fields in  the bosonic sector are  the graviton, $g_{\mu\nu}$, and four complex scalar fields:   three   vector scalars, $z_j$, $j=1,2,3$, and a hypermultiplet scalar, $z$. They parametrize the cosets, $\cals M_V$ and $\cals M_H$, respectively, where
\begin{equation}\label{}
\cals M_V\times \cals M_H\equiv \Bigg[{\rm SU(1,1)\over \rm U(1)}\Bigg]^2\times {\rm SU(1,1)\over  U(1)}\,,
\end{equation}
are  $3+1$ copies of the Poincar\'e disk, with the corresponding K\"ahler potentials: 
\begin{align}\label{KvP}
K_{V} & \eql -\log\big[ (1-|z_1|^2)(1-|z_2|^2)(1-|z_3|^2)\big]\,,\qquad 
K_{H}  \eql -\log(1-|z|^2)\,. 
\end{align}

The bosonic Lagrangian in the $\rm U(1)^3$-invariant sector is 
\begin{equation}\label{4dLagrLor}
e^{-1}\mathcal{L}_\text{bosonic} = \frac{1}{2} R - \sum_{i=1}^3 \frac{\partial_\mu z_i \partial^\mu \bar z_i}{(1-z_i \bar z_i)^2} - \frac{\partial_\mu z \partial^\mu \bar z}{(1-z \bar z)^2} - g^2 \mathcal{P} \,,
\end{equation}
where $e=\sqrt{-\det g_{\mu\nu}}$. The scalar potential, $\cals P$, is given in terms of the  ``holomorphic'' superpotential \cite{Bobev:2018uxk}
\begin{equation}\label{holV}
\mathcal{V} = \frac{|z|^2}{1-|z|^2}(1-z_1)(1-z_2)(1-z_3)+\frac{2}{1-|z|^2}(z_1z_2z_3-1)\,,
\end{equation}
by 
\begin{equation}\label{scpoten}
\mathcal{P} = \frac{1}{2}e^{K_V}\Bigg[\,\sum_{i=1}^3{\nabla_{z_i}\mathcal{V}\nabla_{\bar{z}_i}\overline{\mathcal{V}}\over (1-|z_i|^2)^2}+4{\partial_{z}\mathcal{V}\partial_{\bar{z}}\overline{\mathcal{V}}\over (1-|z|^2)^2}-3\mathcal{V}\overline{\mathcal{V}}\,\Bigg]\;,
\end{equation}
where
\begin{equation}\label{defnabla}
\nabla_{z_i}\mathcal{V} = \partial_{z_i}\mathcal{V}+(\partial_{z_i}K_V)\mathcal{V}\;.
\end{equation}
%

\subsection{Supersymmetry variations}

The supersymmetry variations of  the  gravitini, $\psi_\mu{}^i$,  and the spin-1/2 fields, $\chi^{ijk}$,  of the $\cals N=8$ $d=4$ supergravity are given by \cite{deWit:1982bul} 
\begin{equation}
\delta\psi_\mu{}^i   \eql 2D_\mu\epsilon^i+\sqrt 2 g A_1{}^{ij}\gamma_\mu\epsilon_j\,,\label{spin32}
\end{equation}
\begin{equation}\label{spin12}
\delta\chi^{ijk} \eql -\cals A_\mu{}^{ijkl}\gamma^\mu\epsilon_l-2g A_{2l}{}^{ijk}\epsilon^l\,,
\end{equation}
together with the complex conjugate transformations for the  fields $\psi_{\mu i}$ and $\chi_{ijk}$.\footnote{We recall that in the symmetric gauge for the scalar fields, the conventions of \cite{deWit:1982bul} imply that the complex conjugation amounts to raising/lowering of the $\rm SO(8)$ indices $i,j,k,\ldots =1,\ldots, 8$. It also changes the chirality of the fermion fields.}

The $\rm U(1)^2\subset SU(3)\subset SO(6)\subset SO(8)$ symmetry acts nontrivially on the $\rm SO(8)$ indices $i,j,k=1,\ldots, 6$. The invariant fermion fields are then the chiral gravitini, $\psi_\mu{}^{7,8}$ and $\psi_{\mu\,7,8}$, and the chiral spin-1/2 fields 
\begin{equation}\label{spin12v}
\chi^{127}\,,\quad \chi^{128}\,,\quad \chi^{347}\,,\quad \chi^{348}\,,\quad \chi^{567}\,,\quad \chi^{568}\,,
\end{equation}
\begin{equation}\label{spin12h}
\chi^{135}\eql -\chi^{146}\eql -\chi^{236}\eql -\chi^{245}\,,\qquad  
\chi^{246}\eql -\chi^{136}\eql -\chi^{145}\eql -\chi^{235}\,,
\end{equation}
and their complex conjugates. The supersymmetry parameters in the resulting $\cals N=2$ supergravity are the chiral spinors $\epsilon^{7,8}$ and $\epsilon_{7,8}$, which we relabel henceforth as $\epsilon^{1,2}$ and $\epsilon_{1,2}$, respectively.

By evaluating the scalar tensors $\cals A_{\mu}{}^{ijkl}$, $A_1{}^{ij}$ and $A_{2\,i}{}^{jkl}$ on the scalar fields above, the variations \eqref{spin32} and \eqref{spin12} for the $\rm U(1)^2$-invariant fermion fields can be written down explicitly. 
To simplify the resulting formulae, it is convenient to define the following  auxiliary functions:
\begin{equation}\label{defofW}
\fW\equiv  e^{K_V/2}\,\cals V\,,\qquad \fF_i\equiv -(1-|z_i|^2)\,D_{z_i}\fW\,, \qquad i=1,2,3\,,
\end{equation}
where
\begin{equation}\label{}
D_{z_i}\eql \partial_{z_i}+{1\over 2}\partial_{z_i}K_V\,,
\end{equation}
is the $\cals N=2$ covariant derivative, and
\begin{equation}\label{defofG}
\fG\equiv e^{K_V/2}\big[\,2(z_1z_2z_3-1)+(1-z_1)(1-z_2)(1-z_3)\,\big]\,.
\end{equation}

Then the spin-3/2 variations \eqref{spin32} are given by 
\begin{equation}\label{varr32}
\begin{split}
\delta\psi_\mu{}^7 & \eql 2 \,\nabla_\mu \epsilon^1+\frak B_\mu \epsilon^1-i\,\frak C_\mu \epsilon^2 -{g\over\sqrt 2}\,\overline\fW\,\gamma_\mu\epsilon_1\,,\\
\delta\psi_\mu{}^8 & \eql  2 \,\nabla_\mu \epsilon^2+\frak B_\mu \epsilon^2+i\,\frak C_\mu \epsilon^1-{g\over\sqrt 2}\,\overline\fW\,\gamma_\mu\epsilon_2\,,
\end{split}
\end{equation}
where
\begin{equation}\label{}
\nabla_\mu\eql \partial_\mu+{1\over 4}\omega_\mu{}_{mn}\gamma^{mn}\,,
\end{equation}
is the gravitational covariant derivative and 
\begin{equation}\label{Kgflds}
\frak B_\mu\eql {1\over 2}\sum_j {z_j\partial_\mu \bar z_j-\bar z_j\partial_\mu z_j\over 1-|z_j|^2}\,,\qquad 
\frak C_\mu\eql {1\over 2}\,{z\partial_\mu\bar z-\bar z\partial_\mu z\over 1-|z|^2}\,,
\end{equation}
are the composite K\"ahler gauge fields. 

\noindent
{\it Comment:\/} Note that $\frak C_\mu$ is also the current for the  minimal coupling of the hypermultiplet scalar, $z$,  to  a combination of the  $\rm U(1)^4$ gauge fields. Hence, the consistency of our truncation requires that we set $\frak C_\mu=0$, which implies that the phase of the hypermultiplet scalar, $z$, must be constant.

The spin-1/2 variations for the fields in the vector supermultiplets are given by
\begin{equation}\label{var12}
\begin{split}
\delta\chi^{127}& \eql -\sqrt 2\,{\partial_\mu \bar z_1\over 1-|z_1|^2}\,\gamma^\mu\epsilon_2+g\,\fF_1\,\epsilon^2\,,\qquad 
\delta\chi^{128} \eql \sqrt 2\,{\partial_\mu \bar z_1\over 1-|z_1|^2}\,\gamma^\mu\epsilon_1-g\,\fF_1\,\epsilon^1\,,\\
\delta\chi^{347}& \eql -\sqrt 2\,{\partial_\mu \bar z_2\over 1-|z_2|^2}\,\gamma^\mu\epsilon_2+g\,\fF_2\,\epsilon^2\,,\qquad 
\delta\chi^{348} \eql \sqrt 2\,{\partial_\mu \bar z_2\over 1-|z_2|^2}\,\gamma^\mu\epsilon_1-g\,\fF_2\,\epsilon^1\,,\\
\delta\chi^{567}& \eql -\sqrt 2\,{\partial_\mu \bar z_3\over 1-|z_3|^2}\,\gamma^\mu\epsilon_2+g\,\fF_3\,\epsilon^2\,,\qquad 
\delta\chi^{568} \eql \sqrt 2\,{\partial_\mu \bar z_3\over 1-|z_3|^2}\,\gamma^\mu\epsilon_1-g\,\fF_3\,\epsilon^1\,,\\
\end{split}
\end{equation}
while for those in the hypermultiplet by 
\begin{equation}\label{var12b}
\begin{split}
\delta\chi^{135}   &\eql-\delta\chi^{146}\eql -\delta\chi^{236}\eql -\delta\chi^{245} \\ &\eql
{1\over 2\sqrt 2}\, \left[ {\partial_\mu(\bar z- z) \over 1-|z|^2}\,\gamma^\mu\epsilon_1+ i\, {\partial_\mu(\bar z+ z)\over 1-|z|^2}\,\gamma^\mu\epsilon_2\right]+{g\over 2}\,{\fG\over 1-|z|^2} \left[ (\bar z-z)\epsilon^1+i\,(\bar z+z)\epsilon^2\right]\,,\\[6 pt]
\delta\chi^{246}& \eql -\delta\chi^{136}\eql -\delta\chi^{145}\eql -\delta\chi^{235}\\
& \eql {1\over 2\sqrt 2}\left[i{\partial_\mu(\bar z+z)\over 1-|z|^2}\,\gamma^\mu\epsilon_1 -{\partial_\mu(\bar z -z)\over 1-|z|^2}\,\gamma^\mu\epsilon_2\right]+
{g\over 2}{\fG\over 1-|z|^2}\left[i\,(\bar z+z)\epsilon^1-(\bar z-z)\epsilon^2\right]\,,
\end{split}
\end{equation}
with the complex conjugate transformations for the fields of the opposite chirality.

\section{Euclidean BPS equations}
\label{AppendixB}

The Wick rotation of the supersymmetry variations \eqref{varr32}, \eqref{var12} and \eqref{var12b} can be done in the same way as for the STU-model in \cite{Freedman:2013ryh}. Our goal here is to derive in a somewhat more direct way the BPS equations that result from setting the supersymmetry variations to zero. 

Following \cite{Freedman:2013ryh}, we take the Euclidean metric of the form (cf. \eqref{Ansatz})
\begin{equation}\label{EucMetr}
ds^2\eql L^2\,e^{2A(r)}ds_{S^3}^2+e^{2B(r)}dr^2\,,\qquad ds_{S^3}^2\eql {1\over 4}(\sigma_1^2+\sigma_2^2+\sigma_3^2)\,,
\end{equation}
where $ds_{S^3}^2$ is  the metric on the unit radius $S^3$ with the Maurer-Cartan forms $\sigma_i$, $i=1,2,3$, satisfying $d\sigma_1=\sigma_2\wedge\sigma_3$, etc. A natural set of frames for  the metric \eqref{EucMetr} is then 
\begin{equation}\label{}
e^1\eql {L\over 2} e^A \sigma_1\,,\qquad e^2\eql {L\over 2} e^A \sigma_2\,,\qquad e^3\eql {L\over 2} e^A \sigma_3\,,\qquad 
e^4\eql e^B dr\,.
\end{equation}
As usual, $L=1/\sqrt 2 g$.

Upon Wick rotation, see Section~\ref{sec:Sugra}, the scalar fields $z_i$ and $z$ and their complex conjugates $\bar z_i$ and $\bar z$ become independent complex fields. To emphasize this we denote the latter as $\tilde z_j$ and $\tilde z$, respectively. Replacing $\bar z_i$ and $\bar z$ by $\tilde z_i$ and $\tilde z$ in \eqref{defofW} and \eqref{defofG}, we obtain the  functions   $\fW$, $\fF_j$,  and $\fG$ in the Euclidean regime.  Correspondingly  $\widetilde \fW$, $\widetilde \fF_j$  and $\widetilde \fG$ denote their ``conjugate'' counterparts. They are simply related to the untilded functions by the exchange, $z_i\leftrightarrow \tilde z_i$ and $z\leftrightarrow \tilde z$. For convenience, we have listed these functions explicitly in \eqref{WFG}.

Following \cite{Freedman:2013ryh}, the Wick rotation of  the Dirac matrices amounts to setting 
\begin{equation}\label{}
\gamma^0\quad\longrightarrow\quad -i\,\gamma^4\,,\qquad \gamma^{1,2,3}\quad\longrightarrow\quad \gamma^{1,2,3}\,,
\end{equation}
which implies that 
\begin{equation}\label{gam5cont}
\gamma^5\eql i\,\gamma^0\gamma^1\gamma^2\gamma^3\quad\longrightarrow\quad \gamma^5_E\eql -\gamma^1\gamma^2\gamma^3\gamma^4\,.
\end{equation}
At the same time,    the chirality of the now  independent left- and right-handed complex spinors  remains unchanged  by the rotation. In particular, the supersymmetry parameters now satisfy
\begin{equation}\label{}
\gamma^5_E\epsilon^\alpha \eql \epsilon^\alpha\,,\qquad \gamma^5_E\epsilon_\alpha \eql - \epsilon_\alpha\,,\qquad \alpha=1,2\,.
\end{equation}
Together with \eqref{gam5cont}, this implies
\begin{equation}\label{projijk} 
\gamma^i\gamma^4\epsilon^\alpha\eql \gamma^j\gamma^k\epsilon^\alpha\,,\qquad 
\gamma^i\gamma^4\epsilon_\alpha\eql -\gamma^j\gamma^k\epsilon_\alpha\,, \qquad \alpha=1,2\,,
\end{equation}
where $(ijk)$ is a cyclic permutation of $(123)$.

Let us start with the spin-1/2 variations. Assuming that the scalar fields depend only on the radial coordinate, $r$, the variations in \eqref{var12} read (there is no summation over the repeated index below)
\begin{equation}\label{pairs12}
\begin{split}
\sqrt 2\,{   z_j'\over 1-z_j\tilde z_j}\,e^{-B}\,\gamma^4\epsilon^\alpha-g\,\widetilde \fF_j\,\epsilon_\alpha & \eql 0\,,\\[6 pt]
\sqrt 2\,{  \tilde z_j'\over 1-z_j\tilde z_j}\,e^{-B}\,\gamma^4\epsilon_\alpha-g\,\fF_j\,\epsilon^\alpha & \eql 0\,,\qquad \alpha=1,2\,,\quad j=1,2,3\,.
\end{split}
\end{equation}
We are interested in maximally supersymmetric configurations with non-vanishing Killing spinors  $\epsilon^\alpha$ and $\epsilon_\alpha$.  Thus, if any of the $z_j$ or $\tilde z_j$ is constant, that is $z_j'=0$ or $\tilde z_j'=0$, the corresponding function $\widetilde \fF_j$ or $\fF_j$, respectively,  must vanish. For now, let us assume that the scalars have nontrivial profiles. We will return to the special cases afterwards.

By considering pairs of equations in \eqref{pairs12}, we obtain consistency conditions 
\begin{equation}\label{const1}
\begin{split}
(1-z_i\tilde z_i)\,\widetilde \fF_i\,   z'_j- (1-z_j\tilde z_j)\,\widetilde  \fF_j \,  z'_i & \eql 0\,,\\[6 pt]
 (1-z_i\tilde z_i)\,\widetilde \fF_i\, z'_j- (1-z_j\tilde z_j)\,\widetilde \fF_j \, z'_i& \eql 0\,,
\end{split}
\end{equation}
and 
\begin{equation}\label{const2}
z'_i\tilde z'_j-{1\over 2}g^2 e^{2B} (1-z_i\tilde z_i)(1-z_j\tilde z_j)\,\fF_i\widetilde \fF_j\eql 0\,,
\end{equation}
which must hold for all $i,j=1,2,3$. The conditions \eqref{const1} imply that 
\begin{equation}\label{defofMtM}
M \equiv {1\over 2}\,g\,e^B(1-z_i\tilde z_i)\,{ \widetilde\fF_i\over z_i'}\,,\qquad 
\widetilde M\equiv {1\over 2}\,g\,e^B(1-z_i\tilde z_i)\,{   \fF_i\over\tilde z_i'}\,,
\end{equation}
are the same for all $i$ and, using \eqref{const2}, satisfy 
\begin{equation}\label{mmtilde}
M\widetilde M \eql 1\,.
\end{equation}
This reduces the variations \eqref{pairs12} for generic scalar fields to two projections
\begin{equation}\label{projM} 
\gamma^4\epsilon^\alpha\eql M\epsilon_\alpha\,,\qquad \gamma^4\epsilon_\alpha\eql \widetilde M \epsilon^\alpha\,,
\end{equation}
on the Killing spinors. Finally, using \eqref{mmtilde} we can rewrite \eqref{defofMtM} as the first set of BPS equations
\begin{equation}\label{zjeqs}
z_j'\eql {g\over \sqrt 2}(1-z_j\tilde z_j)\,e^B\,\widetilde M\,\widetilde\fF_j\,,
\qquad 
\tilde z_j'\eql {g\over \sqrt 2}(1-z_j\tilde z_j)\,e^B\,M\,\fF_j\,.
\end{equation}

The remaining spin-1/2 variations \eqref{var12b} yield the following four equations
\begin{equation}\label{newvar12}
\begin{split}
(z'-\tilde z')\gamma^4\epsilon^1-i\,(z'+\tilde z')\gamma^4\epsilon^2+\sqrt 2\,g\,e^B\,\widetilde \fG\,(z-\tilde z)\epsilon_1-i\,\sqrt 2\,g\,e^B\,\widetilde \fG\,(z+\tilde z)\epsilon_2&\eql 0\,,\\[6 pt]
i\,(z'+\tilde z')\gamma^4\epsilon^1+(z'-\tilde z')\gamma^4\epsilon^2 +i\,\sqrt 2\,g\,e^B\,\widetilde \fG\,(z+\tilde z)\epsilon_1+\sqrt 2\,g\,e^B\,\widetilde \fG\,(z-\tilde z)\epsilon_2&\eql 0\,,\\[6 pt]
( z'-\tilde z')\gamma^4\epsilon_1-i \,(z'+\tilde  z')\gamma^4\epsilon_2+\sqrt 2\,g\,\fG\, ( z-\tilde z)\epsilon^1-i\,\sqrt 2\,g\,\fG\,(z+\tilde z)\epsilon^2  & \eql 0\,,\\[6 pt]
i\,( z'+\tilde z')\gamma^4\epsilon_1+(z'-\tilde z')\gamma^4\epsilon_2+i\,\sqrt 2\,g\,\fG\, ( z+\tilde z)\epsilon^1+\sqrt 2\,g\,\fG\,(z-\tilde z)\epsilon^2  & \eql 0\,,
\end{split}
\end{equation} 
which are obviously absent if the hyperscalar vanishes. 
Requiring that there are no further algebraic constraints on the Killing spinors, $\epsilon^\alpha$ and $\epsilon_\alpha$, other than the projections \eqref{projijk} and \eqref{projM}, we find that the following equations must hold:
\begin{equation}\label{zpzjeqs}  
\begin{split}
M\,z'\eql -\sqrt 2\,g\,e^B\,\widetilde \fG\,z\,,\qquad \widetilde M\,z'\eql -\sqrt 2\,g\,e^B\,\fG\,z\,,\\[6 pt]
M\,\tilde z'\eql -\sqrt 2\,g\,e^B\,\widetilde \fG\,\tilde z\,,\qquad \widetilde M\,\tilde z'\eql -\sqrt 2\,g\,,e^B\,\fG\,\tilde z\,.\\
\end{split}
\end{equation}
Using \eqref{mmtilde}, consistency of these equations implies that
\begin{equation}\label{MMGG} 
M \,\fG\eql \widetilde M\,\widetilde\fG\,,
\end{equation}
and we are left with the second set of BPS equations
\begin{equation}\label{zeqs}
{z'}\eql -g\,\sqrt 2\,z\,e^B\, \widetilde M\,\widetilde \fG\,,\qquad 
{\tilde z'}\eql -g\,\sqrt 2\,\tilde z\,e^B\,   M\,  \fG\,.
\end{equation}

Using the projections \eqref{projijk} and \eqref{projM} in the spin-3/2 variations \eqref{varr32} along $S^3$ we find
\begin{equation}\label{vars32sd}
\partial _{\sigma_j}\epsilon^\alpha\eql \Phi\,\gamma^j\epsilon_\alpha\,,\qquad \partial_{\sigma_j}\epsilon_\alpha\eql \widetilde \Phi\,\gamma^j\epsilon^\alpha\,, \qquad j=1,2,3,
\end{equation}
where
\begin{equation}\label{thePhis}
\begin{split}
\Phi & \eql -{1\over 4}\,\Big[L\,e^{A-B}M\,A'-{1\over 2}e^A\,\widetilde \fW+M\,\Big]\,,\qquad 
\widetilde \Phi  \eql -{1\over 4}\,\Big[L\,e^{A-B}\widetilde M\,A'-{1\over 2}e^A\, \fW-\widetilde M\,\Big]\,.
\end{split}
\end{equation}
The derivatives  $\partial_{\sigma_j}$ are along the Killing vector fields dual to the Maurer-Cartan forms $\sigma_j$. Evaluating the commutators 
\begin{equation}\label{commsu2}
[\partial_{\sigma_i},\partial_{\sigma_j}]\eql- \partial_{\sigma_k}\,,
\end{equation}
where $(ijk)$ is a cyclic permutation of $(123)$, on the Killing spinors using  \eqref{vars32sd}, we obtain
\begin{equation}
\begin{split}
 [\partial_{\sigma_i},\partial_{\sigma_j}]\epsilon^\alpha+ \partial_{\sigma_k}\epsilon^\alpha 
 & \eql (-2\,\Phi\,\widetilde \Phi \,M+\Phi)\,\gamma^k\epsilon_\alpha\,,\\[6 pt]
 [\partial_{\sigma_i},\partial_{\sigma_j}]\epsilon_\alpha+ \partial_{\sigma_k}\epsilon_\alpha & \eql 
 (2\,\Phi\,\widetilde \Phi \,\widetilde M+\widetilde \Phi)\,\gamma^k\epsilon^\alpha\,,
\end{split}
\end{equation}
which imply
\begin{equation}\label{Phieqs}
\Phi\,(2\,\widetilde \Phi \,M-1)\eql 0\,,\qquad 
(2\,\Phi\,\widetilde M+1)\,\widetilde \Phi \eql 0\,.
\end{equation}
There are two solutions to these equations that are consistent with \eqref{mmtilde}, the first one obtained by setting
\begin{equation}\label{fstcomb}
\Phi\eql 0\,,\qquad \widetilde \Phi\eql 0\,,
\end{equation}
and the second one by setting
\begin{equation}\label{sndcomb}
2\, \Phi+M\eql 0\,,\qquad 2\,\widetilde\Phi-\widetilde M\eql 0\,.
\end{equation}
Unpacking \eqref{fstcomb} and \eqref{sndcomb} using \eqref{thePhis}, we obtain two sets of BPS equations:
\begin{equation}\label{dAeqsI}
\text{I.}\qquad 
A'\eql g\,\sqrt 2\,\Big[ - e^{-A+B}+{e^B\over 2}\,\widetilde M\,\widetilde\fW\,\Big]\,,\qquad 
A'\eql  g\,\sqrt 2\,\Big[  e^{-A+B}+{e^B\over 2}\,  M\, \fW\,\Big]\,,
\end{equation}
\begin{equation}\label{dAeqsII}
\text{II.}\qquad A'\eql g\,\sqrt 2\,\Big[ - e^{-A+B}+{e^B\over 2}\, M\,\fW\,\Big]\,,\qquad 
 A'\eql  g\,\sqrt 2\,\Big[  e^{-A+B}+{e^B\over 2}\, \widetilde M\, \widetilde\fW\,\Big]\,,
\end{equation}
that are simply related by the exchange of the tilded and untilded fields.

The remaining spin-3/2 variation along the $r$ direction yields the radial dependence of the Killing spinor and does  not give rise to additional BPS equations.  The reader may note that  we did not assume any specific dependence of the Killing spinors on the coordinates along $S^3$ and only used the integrability conditions \eqref{commsu2}, which guarantee the existence of a solution.

To summarize, we have shown that for generic profiles of the scalar fields, the vanishing supersymmetry variations in the Euclidean regime yield first order differential equations \eqref{zjeqs}, \eqref{zeqs} and \eqref{dAeqsI} or \eqref{dAeqsII} for the scalar fields and the metric function together with the algebraic constraints \eqref{mmtilde} and \eqref{MMGG}. 

One can further simplify this system of BPS equations by first solving the two algebraic constraints for $M$ and $\widetilde M$,
\begin{equation}\label{solforM}
M\eql \xi\,{\widetilde \fG^{1/2}\over \fG^{1/2}}\,,\qquad \widetilde M\eql  \xi\,{\fG^{1/2}\over \widetilde \fG^{1/2}}\,,\qquad \xi\eql \pm 1\,.
\end{equation}
Secondly, we have 
\begin{equation}\label{solAdA}
e^A\eql \pm 4\,\xi\,{\fG^{1/2}\,\widetilde \fG^{1/2}\over \fG\,\widetilde\fW-\widetilde\fG\,\fW}\,,
\qquad 
A'   \eql {e^B\over 4L}\,\xi\,{ \fG\,\widetilde \fW+\widetilde \fG\,\fW \over \fG^{1/2}\,\widetilde\fG^{1/2}}  \,, 
\end{equation}
that follow by solving  \eqref{dAeqsI} (top sign) or  \eqref{dAeqsII} (bottom sign).  One is then left with the ``flow'' equations for the scalars that read
\begin{equation}\label{zzjeqs}
z_j'\eql {e^B\over 2L}\,\xi\,(1-z_j\tilde z_j)\,{\fG^{1/2}\over\widetilde\fG^{1/2}}\,\widetilde\fF_j\,,\qquad
\tilde z_j'\eql {e^B\over 2L}\,\xi\,(1-z_j\tilde z_j)\,{\widetilde \fG^{1/2}\over \fG^{1/2}}\,\fF_j\,,
\end{equation}
and
\begin{equation}\label{zzeqs}
{z'\over z}\eql -{e^B\over L}\,\xi \,\fG^{1/2}\,\widetilde \fG^{1/2}\,,\qquad 
{\tilde z'\over \tilde z}\eql -{e^B\over L}\,\xi\,\fG^{1/2}\,\widetilde \fG^{1/2}\,.
\end{equation}
The second equation in \eqref{solAdA} follows then from the first one after using \eqref{zzjeqs} and \eqref{zzeqs}. 

The flow equations \eqref{zzeqs} imply that $z$ and $\tilde z$ are proportional, which implies that the Euclidean continuation of the current $\frak C_\mu$ in \eqref{Kgflds} vanishes. This is consistent with having the $\rm U(1)^4$ gauge fields set to zero.

When the hyperscalars $z$ and $\tilde{z}$ vanish, \eqref{solAdA} and  \eqref{zzjeqs} are equivalent to the BPS equations derived in \cite{Freedman:2013ryh} for the Euclidean STU model. 

Finally, let us consider the special configurations  with constant scalars that were excluded above.\footnote{We omit all possible intermediate solutions with a subset of scalars being constant  as well as solutions with reduced supersymmetry.}  The variations \eqref{pairs12} yield 
\begin{equation}\label{alge1}
\fF_i\eql \widetilde \fF_i\eql 0\,, 
\end{equation}
with no further projection \eqref{projM}. From \eqref{newvar12} we get
\begin{equation}\label{alge2}
\fG\eql \widetilde \fG\eql 0\qquad \text{or}\qquad z\eql\tilde z\eql 0\,,
\end{equation}
while the spin-3/2 variations imply
\begin{equation}\label{Asqeqs}  
(A')^2\eql {1\over L^2}\Big[e^{-2(A-B)}+{1\over 4}e^{2B}\,\fW\,\widetilde\fW\,\Big]\,.
\end{equation}

Note that \eqref{Asqeqs}  also holds for general scalar profiles. It simply follows from \eqref{dAeqsI} or  \eqref{dAeqsII} by eliminating $M$ and $\widetilde M$ using \eqref{mmtilde}.

\section{Euclidean equations of motion}
\label{AppendixC}

The equations of motion  for the metric functions in \eqref{Ansatz} and scalars with radial dependence only can be obtained most efficiently from a Lagrangian for a corresponding one-dimensional system. To do that one combines the bulk action \eqref{bulkS} with the GH-boundary term $S_\text{GH}$ defined in \eqref{counterterms}. After integration by parts of the $A''$ terms from the Ricci scalar for the metric \eqref{Ansatz} and then integration over the sphere, one is left with a one-dimensional action along the radial coordinate,
\begin{equation}\label{1Dact}
S_\text{1D}\eql \textrm{vol}_{S^3} \int dr\,\cals L_\text{1D}\,,
\end{equation}
where $\textrm{vol}_{S^3}$ is the volume of the sphere and the effective one-dimensional Lagrangian is
\begin{equation}\label{1dL}
\mathcal{L}_\text{1D} =  L^3 e^{3A-B} \left[\,-3 (A')^2-{3 \over L^{2}}\, e^{2(B-A)} + \sum_{i=1}^3 \frac{z_i'\zt_i'}{(1-z_i \zt_i)^2} + \frac{z' \zt'}{(1-z \zt)^2} + \frac{e^{2B}}{2L^2}\, \mathcal{P}\, \right] \,.
\end{equation}
Varying \eqref{1Dact} with respect to $B$ gives the Hamiltonian constraint
\begin{equation}\label{Hconst}
3 (A')^2-3 L^{-2} e^{-2(A-B)} - \frac{z_i'\zt_i'}{(1-z_i \zt_i)^2} - \frac{z' \zt'}{(1-z \zt)^2} + \frac{e^{2B}}{2L^2} \,\mathcal{P}  =0\,,
\end{equation}
while varying with respect to $A$, $z_i$, $\zt_i$, $z$, and $\zt$ yields
\begin{equation}\label{EOMs}
\begin{split}
A''-A'B'+L^{-2}e^{-2(A-B)}+\frac{z_i'\zt_i'}{(1-z_i \zt_i)^2} + \frac{z' \zt'}{(1-z \zt)^2} &=0\,, \\
\zt_i''+(3A'-B')\,\zt_i'+\frac{2 z_1 (\zt_i')^2}{1-z_i\zt_i}- \frac{e^{2B}}{2L^2}(1-z_i \zt_i)^2\, \frac{\partial \mathcal{P}}{\partial z_i}&=0 \,,\\
z_i''+(3A'-B')\,z_i'+\frac{2 \zt_1 (z_i')^2}{1-z_i\zt_i}- \frac{e^{2B}}{2L^2}(1-z_i \zt_i)^2 \,\frac{\partial \mathcal{P}}{\partial \zt_i}&=0 \,,\\
\zt''+(3A'-B')\,\zt'+\frac{2 z (\zt')^2}{1-z\zt}- \frac{e^{2B}}{2L^2}(1-z \zt)^2 \,\frac{\partial \mathcal{P}}{\partial z}&=0 \,,\\
z''+(3A'-B')\,z'+\frac{2 \zt (z')^2}{1-z\zt}- \frac{e^{2B}}{2L^2}(1-z \zt)^2\, \frac{\partial \mathcal{P}}{\partial \zt}&=0 \,,
\end{split}
\end{equation}
where we have used \eqref{Hconst} to simplify the first equation. Different gauges for the radial coordinate, $r$, simply amount to different choices for the function $B(r)$ as discussed in Section~\ref{BPSeqs}.
 

\bibliography{mABJMS3}

\providecommand{\href}[2]{#2}\begingroup\raggedright\begin{thebibliography}{10}

\bibitem{Freedman:2013ryh}
D.~Z. Freedman and S.~S. Pufu, {\it {The holography of $F$-maximization}},
  {\em JHEP} {\bf 03} (2014) 135, [\href{http://arxiv.org/abs/1302.7310}{{\tt
  arXiv:1302.7310}}].

\bibitem{Bobev:2013cja}
N.~Bobev, H.~Elvang, D.~Z. Freedman, and S.~S. Pufu, {\it {Holography for $N =
  2^*$ on $S^4$}},  {\em JHEP} {\bf 07} (2014) 001,
  [\href{http://arxiv.org/abs/1311.1508}{{\tt arXiv:1311.1508}}].

\bibitem{Martelli:2011fu}
D.~Martelli, A.~Passias, and J.~Sparks, {\it {The gravity dual of
  supersymmetric gauge theories on a squashed three-sphere}},  {\em Nucl.
  Phys.} {\bf B864} (2012) 840--868,
  [\href{http://arxiv.org/abs/1110.6400}{{\tt arXiv:1110.6400}}].

\bibitem{Martelli:2011fw}
D.~Martelli and J.~Sparks, {\it {The gravity dual of supersymmetric gauge
  theories on a biaxially squashed three-sphere}},  {\em Nucl. Phys.} {\bf
  B866} (2013) 72--85, [\href{http://arxiv.org/abs/1111.6930}{{\tt
  arXiv:1111.6930}}].

\bibitem{Martelli:2012sz}
D.~Martelli, A.~Passias, and J.~Sparks, {\it {The supersymmetric NUTs and bolts
  of holography}},  {\em Nucl. Phys.} {\bf B876} (2013) 810--870,
  [\href{http://arxiv.org/abs/1212.4618}{{\tt arXiv:1212.4618}}].

\bibitem{Farquet:2014kma}
D.~Farquet, J.~Lorenzen, D.~Martelli, and J.~Sparks, {\it {Gravity duals of
  supersymmetric gauge theories on three-manifolds}},  {\em JHEP} {\bf 08}
  (2016) 080, [\href{http://arxiv.org/abs/1404.0268}{{\tt arXiv:1404.0268}}].

\bibitem{Martelli:2013aqa}
D.~Martelli and A.~Passias, {\it {The gravity dual of supersymmetric gauge
  theories on a two-parameter deformed three-sphere}},  {\em Nucl. Phys.} {\bf
  B877} (2013) 51--72, [\href{http://arxiv.org/abs/1306.3893}{{\tt
  arXiv:1306.3893}}].

\bibitem{Bobev:2016nua}
N.~Bobev, H.~Elvang, U.~Kol, T.~Olson, and S.~S. Pufu, {\it {Holography for $
  \mathcal{N} $ = 1$^{*}$ on S$^{4}$}},  {\em JHEP} {\bf 10} (2016) 095,
  [\href{http://arxiv.org/abs/1605.00656}{{\tt arXiv:1605.00656}}].

\bibitem{Gutperle:2018axv}
M.~Gutperle, J.~Kaidi, and H.~Raj, {\it {Mass deformations of 5d SCFTs via
  holography}},  {\em JHEP} {\bf 02} (2018) 165,
  [\href{http://arxiv.org/abs/1801.00730}{{\tt arXiv:1801.00730}}].

\bibitem{Toldo:2017qsh}
C.~Toldo and B.~Willett, {\it {Partition functions on 3d circle bundles and
  their gravity duals}},  {\em JHEP} {\bf 05} (2018) 116,
  [\href{http://arxiv.org/abs/1712.08861}{{\tt arXiv:1712.08861}}].

\bibitem{Bobev:2018ugk}
N.~Bobev, P.~Bomans, and F.~F. Gautason, {\it {Spherical Branes}},  {\em JHEP}
  {\bf 08} (2018) 029, [\href{http://arxiv.org/abs/1805.05338}{{\tt
  arXiv:1805.05338}}].

\bibitem{Aharony:2008ug}
O.~Aharony, O.~Bergman, D.~L. Jafferis, and J.~Maldacena, {\it {N=6
  superconformal Chern-Simons-matter theories, M2-branes and their gravity
  duals}},  {\em JHEP} {\bf 10} (2008) 091,
  [\href{http://arxiv.org/abs/0806.1218}{{\tt arXiv:0806.1218}}].

\bibitem{Kapustin:2009kz}
A.~Kapustin, B.~Willett, and I.~Yaakov, {\it {Exact Results for Wilson Loops in
  Superconformal Chern-Simons Theories with Matter}},  {\em JHEP} {\bf 03}
  (2010) 089, [\href{http://arxiv.org/abs/0909.4559}{{\tt arXiv:0909.4559}}].

\bibitem{Jafferis:2011zi}
D.~L. Jafferis, I.~R. Klebanov, S.~S. Pufu, and B.~R. Safdi, {\it {Towards the
  F-Theorem: N=2 Field Theories on the Three-Sphere}},  {\em JHEP} {\bf 06}
  (2011) 102, [\href{http://arxiv.org/abs/1103.1181}{{\tt arXiv:1103.1181}}].

\bibitem{Jafferis:2010un}
D.~L. Jafferis, {\it {The Exact Superconformal R-Symmetry Extremizes Z}},  {\em
  JHEP} {\bf 05} (2012) 159, [\href{http://arxiv.org/abs/1012.3210}{{\tt
  arXiv:1012.3210}}].

\bibitem{Closset:2012vg}
C.~Closset, T.~T. Dumitrescu, G.~Festuccia, Z.~Komargodski, and N.~Seiberg,
  {\it {Contact Terms, Unitarity, and F-Maximization in Three-Dimensional
  Superconformal Theories}},  {\em JHEP} {\bf 10} (2012) 053,
  [\href{http://arxiv.org/abs/1205.4142}{{\tt arXiv:1205.4142}}].

\bibitem{Pufu:2016zxm}
S.~S. Pufu, {\it {The F-Theorem and F-Maximization}},  {\em J. Phys.} {\bf A50}
  (2017), no.~44 443008, [\href{http://arxiv.org/abs/1608.02960}{{\tt
  arXiv:1608.02960}}].

\bibitem{Benna:2008zy}
M.~Benna, I.~Klebanov, T.~Klose, and M.~Smedback, {\it {Superconformal
  Chern-Simons Theories and AdS(4)/CFT(3) Correspondence}},  {\em JHEP} {\bf
  09} (2008) 072, [\href{http://arxiv.org/abs/0806.1519}{{\tt
  arXiv:0806.1519}}].

\bibitem{Klebanov:2008vq}
I.~Klebanov, T.~Klose, and A.~Murugan, {\it {AdS(4)/CFT(3) Squashed, Stretched
  and Warped}},  {\em JHEP} {\bf 03} (2009) 140,
  [\href{http://arxiv.org/abs/0809.3773}{{\tt arXiv:0809.3773}}].

\bibitem{Warner:1983vz}
N.~P. Warner, {\it {Some New Extrema of the Scalar Potential of Gauged $N=8$
  Supergravity}},  {\em Phys. Lett.} {\bf B128} (1983) 169--173.

\bibitem{Corrado:2001nv}
R.~Corrado, K.~Pilch, and N.~P. Warner, {\it {An N=2 supersymmetric membrane
  flow}},  {\em Nucl. Phys.} {\bf B629} (2002) 74--96,
  [\href{http://arxiv.org/abs/hep-th/0107220}{{\tt hep-th/0107220}}].

\bibitem{Ahn:2000mf}
C.-h. Ahn and K.~Woo, {\it {Supersymmetric domain wall and RG flow from
  4-dimensional gauged N=8 supergravity}},  {\em Nucl. Phys.} {\bf B599} (2001)
  83--118, [\href{http://arxiv.org/abs/hep-th/0011121}{{\tt hep-th/0011121}}].

\bibitem{Ahn:2000aq}
C.-h. Ahn and J.~Paeng, {\it {Three-dimensional SCFTs, supersymmetric domain
  wall and renormalization group flow}},  {\em Nucl. Phys.} {\bf B595} (2001)
  119--137, [\href{http://arxiv.org/abs/hep-th/0008065}{{\tt hep-th/0008065}}].

\bibitem{Bobev:2009ms}
N.~Bobev, N.~Halmagyi, K.~Pilch, and N.~P. Warner, {\it {Holographic, N=1
  Supersymmetric RG Flows on M2 Branes}},  {\em JHEP} {\bf 09} (2009) 043,
  [\href{http://arxiv.org/abs/0901.2736}{{\tt arXiv:0901.2736}}].

\bibitem{Bobev:2018uxk}
N.~Bobev, V.~S. Min, and K.~Pilch, {\it {Mass-deformed ABJM and black holes in
  AdS$_{4}$}},  {\em JHEP} {\bf 03} (2018) 050,
  [\href{http://arxiv.org/abs/1801.03135}{{\tt arXiv:1801.03135}}].

\bibitem{Drukker:2010nc}
N.~Drukker, M.~Marino, and P.~Putrov, {\it {From weak to strong coupling in
  ABJM theory}},  {\em Commun. Math. Phys.} {\bf 306} (2011) 511--563,
  [\href{http://arxiv.org/abs/1007.3837}{{\tt arXiv:1007.3837}}].

\bibitem{deWit:1982bul}
B.~de~Wit and H.~Nicolai, {\it {N=8 Supergravity}},  {\em Nucl. Phys.} {\bf
  B208} (1982) 323.

\bibitem{deWit:1986oxb}
B.~de~Wit and H.~Nicolai, {\it {The Consistency of the S**7 Truncation in D=11
  Supergravity}},  {\em Nucl. Phys.} {\bf B281} (1987) 211--240.

\bibitem{Nicolai:2011cy}
H.~Nicolai and K.~Pilch, {\it {Consistent Truncation of d = 11 Supergravity on
  AdS$_4 \times S^7$}},  {\em JHEP} {\bf 03} (2012) 099,
  [\href{http://arxiv.org/abs/1112.6131}{{\tt arXiv:1112.6131}}].

\bibitem{Skenderis:2002wp}
K.~Skenderis, {\it {Lecture notes on holographic renormalization}},  {\em
  Class. Quant. Grav.} {\bf 19} (2002) 5849--5876,
  [\href{http://arxiv.org/abs/hep-th/0209067}{{\tt hep-th/0209067}}].

\bibitem{BREITENLOHNER1982249}
P.~Breitenlohner and D.~Z. Freedman, {\it Stability in gauged extended
  supergravity},  {\em Annals of Physics} {\bf 144} (1982), no.~2 249 -- 281.

\bibitem{Klebanov:1999tb}
I.~R. Klebanov and E.~Witten, {\it {AdS / CFT correspondence and symmetry
  breaking}},  {\em Nucl. Phys.} {\bf B556} (1999) 89--114,
  [\href{http://arxiv.org/abs/hep-th/9905104}{{\tt hep-th/9905104}}].

\bibitem{Freedman:2016yue}
D.~Z. Freedman, K.~Pilch, S.~S. Pufu, and N.~P. Warner, {\it {Boundary Terms
  and Three-Point Functions: An AdS/CFT Puzzle Resolved}},  {\em JHEP} {\bf 06}
  (2017) 053, [\href{http://arxiv.org/abs/1611.01888}{{\tt arXiv:1611.01888}}].

\bibitem{Fukuhara:1961a}
M.~Fukuhara, T.~Kimura, and C.~Matsuda, {\em Équations différentielles
  ordinaires du premier ordre dans le champ complexe}.
\newblock [Tokyo] Mathematical Society of Japan, 1961.

\bibitem{Nosaka:2015iiw}
T.~Nosaka, {\it {Instanton effects in ABJM theory with general R-charge
  assignments}},  {\em JHEP} {\bf 03} (2016) 059,
  [\href{http://arxiv.org/abs/1512.02862}{{\tt arXiv:1512.02862}}].

\bibitem{Bobev:2018hbq}
N.~Bobev, F.~F. Gautason, and J.~Van~Muiden, {\it {Precision Holography for
  $\mathcal{N}=2^{*}$ on $S^4$ from type IIB Supergravity}},  {\em JHEP} {\bf
  04} (2018) 148, [\href{http://arxiv.org/abs/1802.09539}{{\tt
  arXiv:1802.09539}}].

\bibitem{Bhattacharyya:2012ye}
S.~Bhattacharyya, A.~Grassi, M.~Marino, and A.~Sen, {\it {A One-Loop Test of
  Quantum Supergravity}},  {\em Class. Quant. Grav.} {\bf 31} (2014) 015012,
  [\href{http://arxiv.org/abs/1210.6057}{{\tt arXiv:1210.6057}}].

\bibitem{Guarino:2015vca}
A.~Guarino and O.~Varela, {\it {Consistent $ \mathcal{N}=8 $ truncation of
  massive IIA on S$^{6}$}},  {\em JHEP} {\bf 12} (2015) 020,
  [\href{http://arxiv.org/abs/1509.02526}{{\tt arXiv:1509.02526}}].

\bibitem{Kim:2018sdw}
H.~Kim, N.~Kim, and M.~Suh, {\it {On the U(1)$^{2}$-Invariant Sector of Dyonic
  Maximal Supergravity}},  {\em J. Korean Phys. Soc.} {\bf 73} (2018), no.~3
  249--258, [\href{http://arxiv.org/abs/1801.01286}{{\tt arXiv:1801.01286}}].

\bibitem{Guarino:2015jca}
A.~Guarino, D.~L. Jafferis, and O.~Varela, {\it {String Theory Origin of Dyonic
  N=8 Supergravity and Its Chern-Simons Duals}},  {\em Phys. Rev. Lett.} {\bf
  115} (2015), no.~9 091601, [\href{http://arxiv.org/abs/1504.08009}{{\tt
  arXiv:1504.08009}}].

\bibitem{Azzurli:2017kxo}
F.~Azzurli, N.~Bobev, P.~M. Crichigno, V.~S. Min, and A.~Zaffaroni, {\it {A
  universal counting of black hole microstates in AdS$_{4}$}},  {\em JHEP} {\bf
  02} (2018) 054, [\href{http://arxiv.org/abs/1707.04257}{{\tt
  arXiv:1707.04257}}].

\bibitem{Duff:1999gh}
M.~J. Duff and J.~T. Liu, {\it {Anti-de Sitter black holes in gauged N = 8
  supergravity}},  {\em Nucl. Phys.} {\bf B554} (1999) 237--253,
  [\href{http://arxiv.org/abs/hep-th/9901149}{{\tt hep-th/9901149}}].

\bibitem{Cvetic:1999xp}
M.~Cvetic, M.~J. Duff, P.~Hoxha, J.~T. Liu, H.~Lu, J.~X. Lu,
  R.~Martinez-Acosta, C.~N. Pope, H.~Sati, and T.~A. Tran, {\it {Embedding AdS
  black holes in ten-dimensions and eleven-dimensions}},  {\em Nucl. Phys.}
  {\bf B558} (1999) 96--126, [\href{http://arxiv.org/abs/hep-th/9903214}{{\tt
  hep-th/9903214}}].

\bibitem{Behrndt:1996hu}
K.~Behrndt, R.~Kallosh, J.~Rahmfeld, M.~Shmakova, and W.~K. Wong, {\it {STU
  black holes and string triality}},  {\em Phys. Rev.} {\bf D54} (1996)
  6293--6301, [\href{http://arxiv.org/abs/hep-th/9608059}{{\tt
  hep-th/9608059}}].

\bibitem{BrittoPacumio:1999sn}
R.~Britto-Pacumio, A.~Strominger, and A.~Volovich, {\it {Holography for coset
  spaces}},  {\em JHEP} {\bf 11} (1999) 013,
  [\href{http://arxiv.org/abs/hep-th/9905211}{{\tt hep-th/9905211}}].

\end{thebibliography}\endgroup
\bibliographystyle{JHEP}

\end{document}